\documentclass[twocolumn,aps,prl,superscriptaddress,floatfix,longbibliography]{revtex4-1}  

\usepackage[draft]{changes}
\usepackage[normalem]{ulem}

\usepackage{graphicx}
\usepackage{amssymb} 
\usepackage{dcolumn}
\usepackage{bm}
\usepackage[mathlines]{lineno}
\usepackage[colorlinks,linkcolor=blue,anchorcolor=blue,citecolor=blue,urlcolor=blue]{hyperref}
\usepackage{multirow}
\usepackage{color}
\usepackage{amsmath}

\makeatletter
\renewcommand*{\@fnsymbol}[1]{\ensuremath{\ifcase#1\or \dagger\or *\or \ddagger\or
\mathsection\or \mathparagraph\or \|\or **\or \dagger\dagger \or
\ddagger\ddagger \else\@ctrerr\fi}} \makeatother

\usepackage{color}%

\usepackage[draft]{changes}
\usepackage[normalem]{ulem}
\definechangesauthor[name={gao}, color=blue]{gao}

\usepackage{color}%

\begin{document}
%
\title{Ambient-pressure superconductivity above 22 K in hole-doped YB$_2$}

%

%

\author{Xuejie Li}
\affiliation{State Key Laboratory for Mechanical Behavior of Materials, School of Materials Science and Engineering,
Xi’an Jiaotong University, Xi’an 710049, China}

\author{Wenbo Zhao}
\affiliation{Key Laboratory of Material Simulation Methods and Software of Ministry of Education, College of Physics, Jilin University, Changchun 130012, China}

\author{Yuzhou Hao}
\affiliation{State Key Laboratory for Mechanical Behavior of Materials, School of Materials Science and Engineering,
Xi’an Jiaotong University, Xi’an 710049, China}

\author{Xiaoying Wang}
\affiliation{State Key Laboratory for Mechanical Behavior of Materials, School of Materials Science and Engineering,
Xi’an Jiaotong University, Xi’an 710049, China}

                



\author{Zhibin Gao}
\email[E-mail: ]{zhibin.gao@xjtu.edu.cn}
\affiliation{State Key Laboratory for Mechanical Behavior of Materials, School of Materials Science and Engineering,
Xi’an Jiaotong University, Xi’an 710049, China}

\author{Xiangdong Ding}
\affiliation{State Key Laboratory for Mechanical Behavior of Materials, School of Materials Science and Engineering,
Xi’an Jiaotong University, Xi’an 710049, China}

\date{\today}
\begin{abstract}
%

Recent studies of hydrogen-dominant (superhydride) materials, such as LaH$_{10}$ have led to putative discoveries of near-room temperature superconductivity at high pressures, with a superconducting transition temperature ($T_c$) of 250 K observed at 170 GPa. While these findings are promising, achieving such high superconductivity requires challenging experimental conditions typically exceeding 100 GPa.
In this study, we utilize first-principles calculations and Migdal-Eliashberg theory to examine the superconducting properties of the stable boron-based compound YB$_2$ at atmospheric pressure, where yttrium and boron atoms form a layered structure. 
Our results indicate that YB$_2$ exhibits a $T_c$ of 2.14 K at 0 GPa. We find that doping with additional electrons (0 to 0.3) leads to a monotonic decrease in $T_c$ as the electron concentration increases. Conversely, introducing holes significantly enhances $T_c$, raising it to 22.83 K. 
Although our findings do not surpass the superconducting temperature of the well-known MgB$_2$, our doping strategy highlights a method for tuning electron-phonon coupling strength in metal borides. This insight could be valuable for future experimental applications. Overall, this study not only deepens our understanding of YB$_2$ superconducting properties but also contributes to the ongoing search for high-temperature superconductors.

\end{abstract}



\maketitle



\section{I. Introduction}

Superconductivity is the complete absence of electrical resistance and is observed in many materials when they are cooled below their superconducting transition temperature ($T_c$). In the Bardeen-Cooper–Schrieffer (BCS) theory of (``conventional'') superconductivity, this occurs when electrons overcome their mutual electrical repulsion and form ``Cooper pairs'' that then travel unheeded through the material as a supercurrent. Low-temperature~\cite{PhysRevLett.9.489,PhysRevB.34.4552,matthias1968superconductivity} and high-temperature superconductors are the two categories of superconducting materials. Although the most common among high-temperature superconductors, such as LaH$_{10}$, are currently hydrogen-based compounds~\cite{drozdov2019superconductivity,cross2024high,chen2024synthesis}, their use is severely limited by the fact that they require extremely high pressure to achieve superconductivity.

The discovery of magnesium diboride, with a superconducting transition temperature of 39 K at ambient pressure~\cite{nagamatsu2001superconductivity}, reignited interest in metal boride superconductors with similar structures. 
However, the $T_c$ of most B-related AlB$_2$-type structures remain below 10 K at 0 GPa, such as NbB$_2$ (0.62 K)~\cite{LEYAROVSKA1979249}, ScB$_2$ (1.5 K)~\cite{10.1063/1.4816117} and MoB$_{2.5}$(8.1 K)~\cite{doi:10.1073/pnas.67.1.313}. 
Elemental doping, such as in Nb$_{0.95}$Y$_{0.05}$B$_{2.5}$ (9.3 K) and Mo$_{0.85}$Zr$_{0.15}$B$_{2.5}$ (11.2 K)~\cite{doi:10.1073/pnas.67.1.313}, is commonly employed to enhance $T_c$, though the impact is often limited. 

Another approach involves pressure, which can raise $T_c$, as seen in $\alpha$-MoB$_{2}$, which achieves a $T_c$ of 37 K at 90 GPa~\cite{PhysRevB.106.064507}. However, this method presents challenges in practical application. Additionally, magnesium diboride analogues with even higher transition temperatures have been identified, with bulk CaB$_2$ being a notable example, having an estimated $T_c$ of around 47 K at 0 GPa~\cite{PhysRevB.80.064503}, though it has yet to be synthesized.

Although the $T_c$ of most B-series compounds has been either calculated or experimentally verified, YB$_2$ has received relatively little attention. As a result, there is a growing belief that this compound lacks superconductivity. Contributing factors may include low electron concentration~\cite{doi:10.1073/pnas.67.1.313}, phonon properties near the Fermi surface and $\Gamma$ point due to a minor hole concentration~\cite{PhysRevB.64.020502}, and high ionicity in Y-B bonds~\cite{XLChen_2001}. 
However, it is noteworthy that similar B-based binary borides, such as ScB$_2$, have been studied within group \uppercase\expandafter{\romannumeral3}$_B$ metal diborides for their superconducting properties~\cite{10.1063/1.4816117}, suggesting that YB$_2$ has the potential to develop superconducting behavior.


According to the BCS theory, the high superconducting $T_c$ of MgB$_2$ primarily arises from its high Debye temperature and strong electron-phonon coupling (EPC). Studies on the electronic structure of MgB$_2$ and related binary borides suggest that the metallic B layers play a crucial role in its superconductivity, particularly due to the presence of $p_{x,y}$-band holes at the $\Gamma$ point. Consequently, when investigating the superconducting properties of MgB$_2$-type metal boride superconductors, it is essential to consider the effects of adding holes and electrons on the superconducting gaps~\cite{li2024tunable,rudenko2024strong,choi2024unified}.


In this work, we investigate the effects of varying effective Coulomb pseudopotential parameters ($\mu^*$), as well as the addition of holes and electrons, on superconducting gaps and $T_c$ of YB$_2$ in detail. Two different software programs are used for this analysis, one based on the isotropic Migdal-Eliashberg equations~\cite{FLORESLIVAS20201,10.1063/5.0077748} and the other on the McMillan-Allen-Dynes formula~\cite{PhysRevB.12.905}. Notably, the $T_c$ of YB$_2$ without any modifications is approximately 2.14 K when $\mu^*$ is set to 0.1 at 0 GPa. However, with increasing hole concentrations from 0 to 0.8, the $T_c$ of YB$_2$ rises gradually at 0.7 to 22.83 K at 0 GPa, after which it falls sharply. In contrast, the addition of electrons leads to a consistent decrease in $T_c$. This not only shows that YB$_2$ is superconductor and that its $T_c$ can be greatly increased by adding a specific concentration of holes, but it also offers fresh insights into the research of B-system compounds, a class of superconductors that operate at atmospheric pressure.

\section{II. COMPUTATIONAL METHODS}
To investigate the structural and electronic properties of YB$_2$ at different hole and electron concentrations, we employed the Vienna \textit{Ab initio} Simulation Package (VASP)~\cite{KRESSE199615,PhysRevB.54.11169} using the Perdew-Burke-Ernzerhof (PBE)~\cite{PhysRevLett.77.3865} Generalized Gradient Approximation (GGA)~\cite{PhysRevB.50.4954,PhysRevB.73.235116} for the DFT calculations. The bare ion Coulomb potential was treated in the projector augmented wave (PAW)~\cite{KRESSE199615,PhysRevB.48.13115,PhysRevB.50.17953} framework. A plane-wave basis set with an energy cutoff of 420 eV and 13 $\times$ 13 $\times$ 13 \textit{k}-point grid were used for the electronic self-consistent calculations. The structures were fully optimized until the maximum energy and force converged to less than 10$^{-6}$ eV/\AA ~ and 1 meV/\AA, respectively. %


The phonon frequencies and electron-phonon coupling (EPC) were calculated using the Quantum-ESPRESSO (QE) package~\cite{Giannozzi_2009} within the framework of density functional perturbation theory (DFPT). Optimized norm-conserving Vanderbilt (ONCV) pseudopotentials~\cite{SCHLIPF201536} were employed, with a kinetic energy cutoff of 60 Ry and a charge density cutoff of 480 Ry. Self-consistent electron density and EPC calculations were performed using a 24 $\times$ 24 $\times$ 24 \textit{k}-point meshes and a 6 $\times$ 6 $\times$ 6 \textit{q}-point meshes. The superconducting critical temperature ($T_c$), phonon spectrum, electronic bands, and density of states (DOS) of YB$_2$ with varying electron and hole concentrations were also computed using a 3 $\times$ 3 $\times$ 3 \textit{q}-point grid, while retaining the same \textit{k}-point grid as used for pure YB$_2$. Additionally, the isotropic Migdal-Eliashberg equations~\cite{FLORESLIVAS20201,10.1063/5.0077748} were solved using the ELK code~\cite{elk} in our study. 


The superconducting critical temperature ($T_c$) of YB$_2$, with effective Coulomb pseudopotential parameters ($\mu^*$) ranging from 0.05 to 0.13 and under different concentrations of electron and hole doping, was calculated using the isotropic Migdal-Eliashberg equations are as follows:

\begin{eqnarray}
\begin{split}
\label{eqn1}	 
Z(i\omega_n) =  1 + \frac{\pi T}{\omega_n} \sum_{n'} \frac{\omega_{n'}}{\sqrt{\omega_{n}^2 +\Delta^2 (i\omega_{n})}} \times \lambda(n-n'),
\end{split}
\end{eqnarray}
%

%
%

\begin{align}
\label{eqn2}
Z(i\omega_n)\Delta(i\omega_n) = \pi T \sum_{n'} \int d\xi\,
&\frac{\Delta(i\omega_{n'})}{\sqrt{\omega_{n'}^2 + \Delta^2(i\omega_{n'})}} \times \\
&\left[\lambda(n-n') - \mu^*\right], \nonumber
\end{align}
%

%
%
%
where $\omega_{n'}$ are the fermionic Matsubara frequencies and $\Delta(i\omega_n)$ is a superconducting gap, with renormalization function $Z(i\omega_n)$. Then the Eliashberg spectral function is defined as:

\begin{eqnarray}
\label{eqn3}	 
\alpha^2 F(\omega) = \frac{1}{2 \pi N(0)} \sum_{qj} \frac{\gamma_{qj}}{\omega_{qj}} \delta(\hbar \omega - \hbar \omega_{qj}),
\end{eqnarray}
where $\gamma_{qj}$ stands for phonon linewidth, and $\omega_{qj}$ represents frequency with a phonon \textit{j} with a wave-vector \textbf{q}

Allen and Dynes performed additional adjustments after McMillan's numerical analysis of the Eliashberg equation for numerous systems, yielding the following the McMillan-Allen-Dynes formula~\cite{PhysRevB.12.905},

\begin{eqnarray}
\label{eqn4}	 
{T}_{c} = \frac{ \omega_{\rm log}}{1.2} \exp \left[-\frac{1.04(1+\lambda)}{\lambda-\mu^*(1+0.62\lambda)}\right],
\end{eqnarray}
where $\mu^*$ is the effective Coulomb pseudopotential and can be set at a typical value of 0.1. The integrated electron-phonon coupling constant is, 
%
%
\begin{eqnarray}
\label{eqn5}	 
\lambda(\omega)= 2 \int_0^\omega \frac{\alpha^2 F(\omega)}{\omega} d\omega,
\end{eqnarray}
in which the electron-phonon coupling constant $\lambda$ used in Eq.~(\ref{eqn1}) is $\lambda(\omega\rm_{max})$, where $\omega\rm_{max}$ is the maximum of the phonon frequency. Besides, the logarithmically averaged characteristic phonon frequency $\omega\rm_{log}$ can be written as,
%
\begin{eqnarray}
\label{eqn6}	 
\omega_{\rm log} = \exp \left[\frac{2}{\lambda} \int \frac{d\omega}{\omega} \alpha^2 F(\omega) \ln(\omega) \right],
\end{eqnarray}
%


In conclusion, the isotropic Migdal-Eliashberg equations is the source of the reduced McMillan-Allen-Dynes formula. By adding certain approximations, such as a simplified treatment of the phonon spectrum, it lowers the computing cost. Additionally, the two formulas have different areas of application. The isotropic Migdal-Eliashberg equations can precisely describe the microscopic features of superconductors and is relevant to highly coupled superconductors. Nonetheless, the McMillan-Allen-Dynes equations is primarily utilized for quick $T_c$ estimate and works well for superconductors with modest coupling strengths.

\begin{figure*}
\includegraphics[width=2.0\columnwidth]{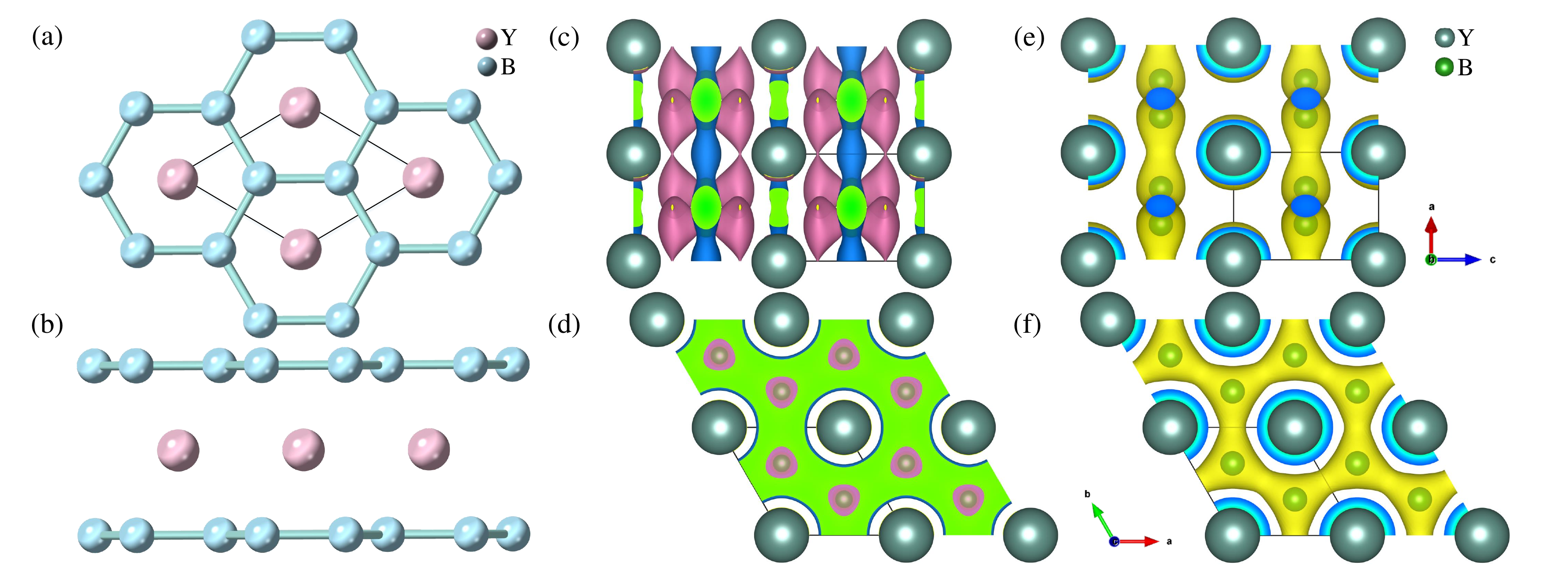}
\caption{%
%
The structural and electronic properties of YB$_2$: Top (a) and side (b) views of the atomic structure, with boron (B) represented by blue atoms and yttrium (Y) by pink atoms. For the side and top views, the charge density difference maps are displayed in (c) and (d), respectively, with purple denoting positive values and blue denoting negative values. Furthermore, (e) and (f) show the wave functions of YB$_2$, where boron (B) is represented by small atoms and yttrium (Y) by large atoms.
\label{fig1}}
\end{figure*}

\begin{figure}
\includegraphics[width=1.0\columnwidth]{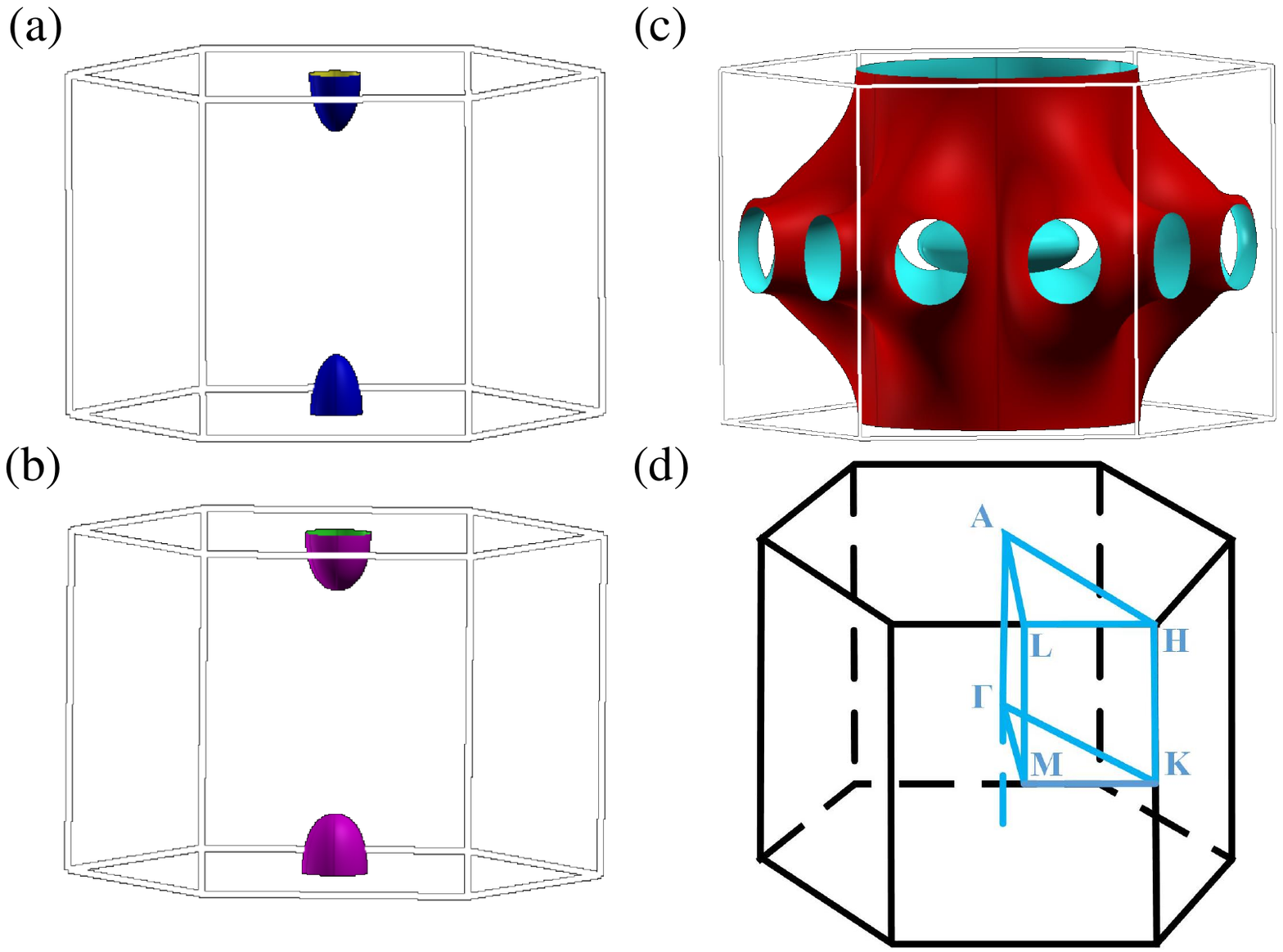}
\caption{%
%
The calculated Fermi surface of YB$_2$ around the A symmetry point is shown for the 1st energy band (a), the 2nd energy band (b), and the 3rd energy band (c). (d) the first Brillouin zone of YB$_2$, with the $k$-path marked by blue lines, consisting of high-symmetry points $\Gamma$, K, M, A, H, and L.
\label{fig2}}
\end{figure}

\begin{figure}
\includegraphics[width=1.0\columnwidth]{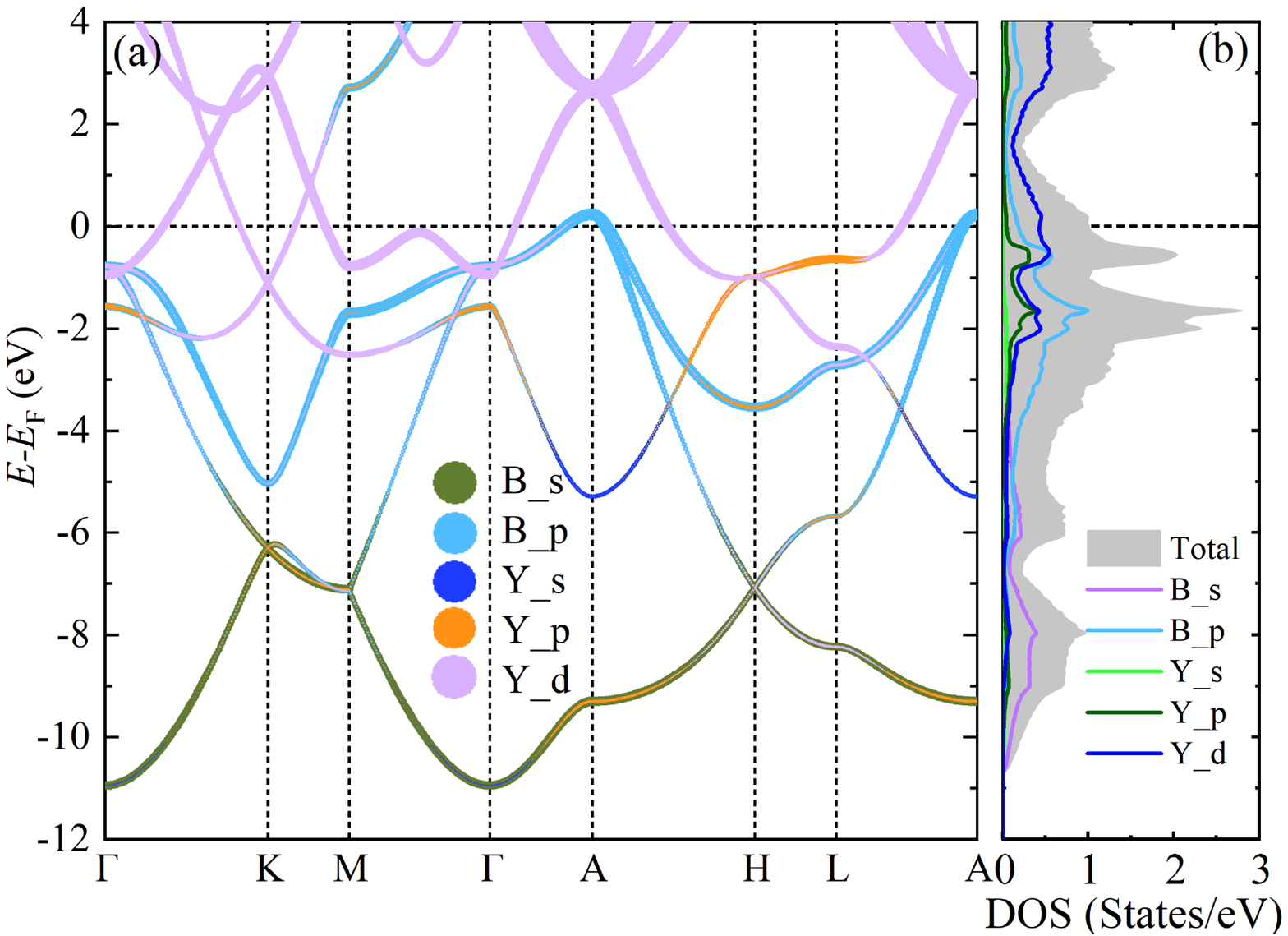}
\caption{
%
The energy band structure and DOS of YB$_2$ are depicted in (a) and (b). In (a), the energy band diagram of YB$_2$ shows the contribution of different orbitals, where green, light blue, dark blue, orange, and purple dots represent the proportion of B-$s$, B-$p$, Y-$s$, Y-$p$, and Y-$d$ orbitals, respectively. In (b), the total DOS of YB$_2$ is shown in shades of gray, with contributions from B-$s$ orbitals in purple, B-$p$ orbitals in light blue, Y-$s$ orbitals in light green, Y-$p$ orbitals in dark green, and Y-$d$ orbitals in dark blue.
\label{fig3}}
\end{figure}

\begin{figure}
\includegraphics[width=1.0\columnwidth]{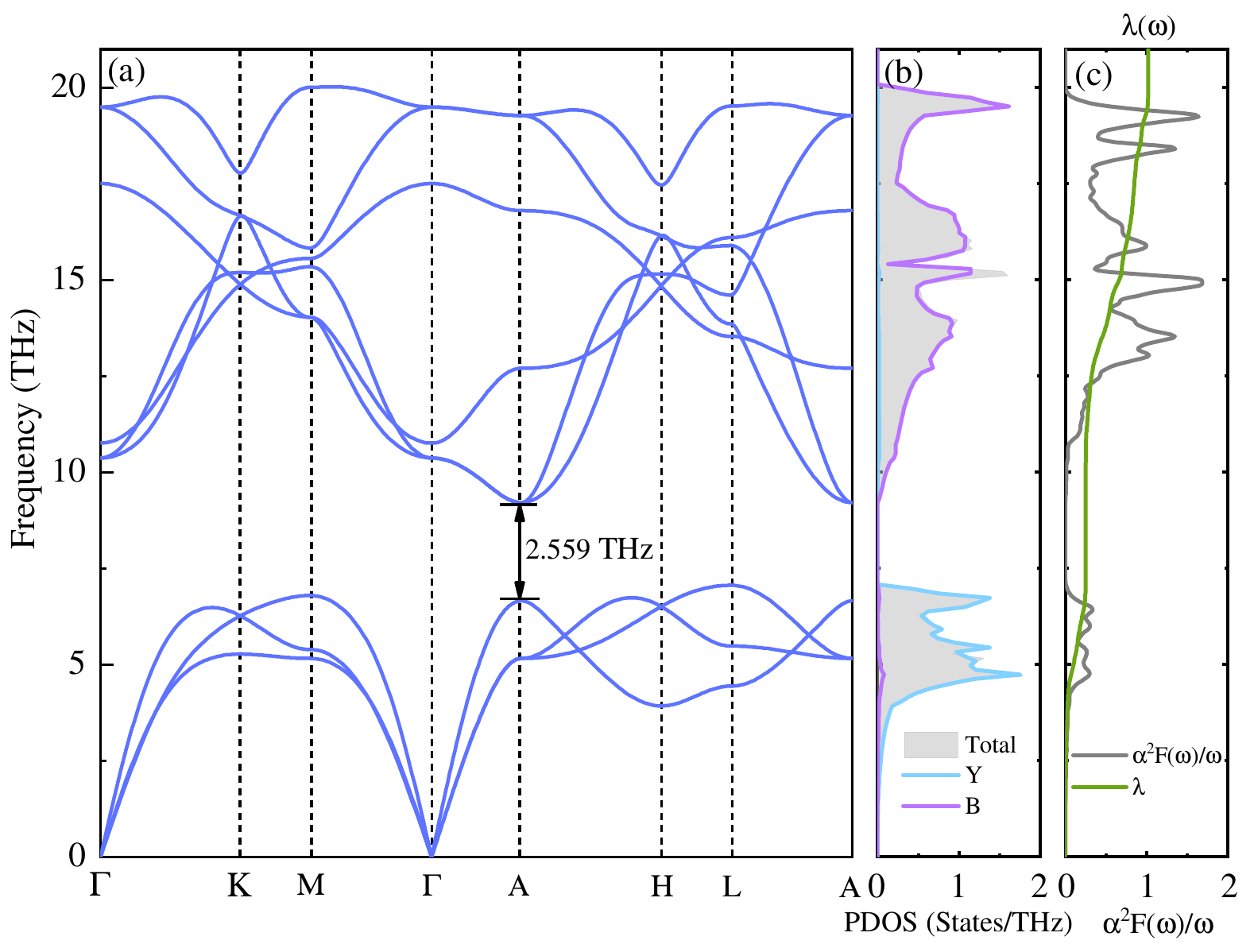}
\caption{
%
The phonon spectrum, the projected phonon density of states (PDOS), the Eliashberg spectral function $\alpha^2$F($\omega$), and the electron-phonon coupling 
$\lambda$($\omega$) of YB$_2$ at ambient pressure are shown. (a) The phonon dispersion diagram is represented by dark blue lines. (b) The total PDOS of YB$_2$ is shaded in light gray, with contributions from Y shown in light blue lines and B in purple lines. (c) The Eliashberg spectral function $\alpha^2$F ($\omega$) is depicted in a dark gray line, alongside the electron-phonon coupling integral $\lambda$($\omega$) in green, for YB$_2$ at ambient pressure.
\label{fig4}}
\end{figure}

\begin{figure*}
\includegraphics[width=2.0\columnwidth]{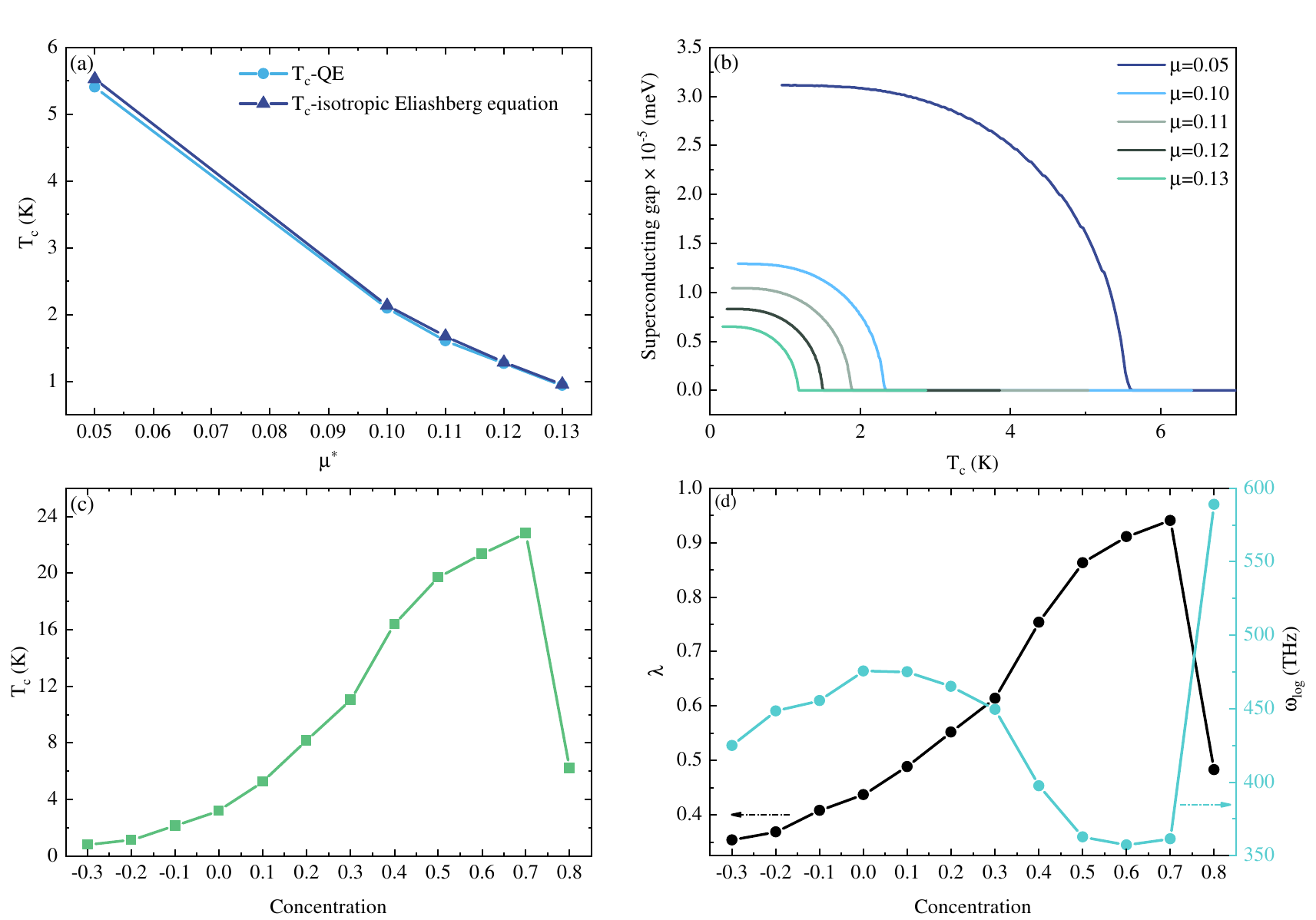}
\caption{
The superconducting properties of YB$_2$. (a) The light blue line stands for the result of $T_c$ in the range of 0.05 to 0.13 by using QE package, while the dark blue line represent the $T_c$ using isotropic Migdal-Eliashberg equation. (b) The isotropic superconducting gap of YB$_2$ at ambient pressure with different $\mu^*$, where $\mu^*$ is equal to 0.05, 0.10, 0.11, 0.12, and 0.13 represented by light-green, dark-green, grayish green, light-blue, and dark-blue lines, respectively. (c) The green line stands for the result of $T_c$ in different concentration of added holes or electrons by using QE package, where positive value mean adding holes, while negative one mean adding electrons. (d) The trends of $\lambda$ and $\omega\rm_{log}$ of YB$_2$ following the addition of various electron or hole concentrations are shown by the aqua blue and black lines, respectively.
%
%
\label{fig5}}
\end{figure*}

\begin{figure*}
\includegraphics[width=2.0\columnwidth]{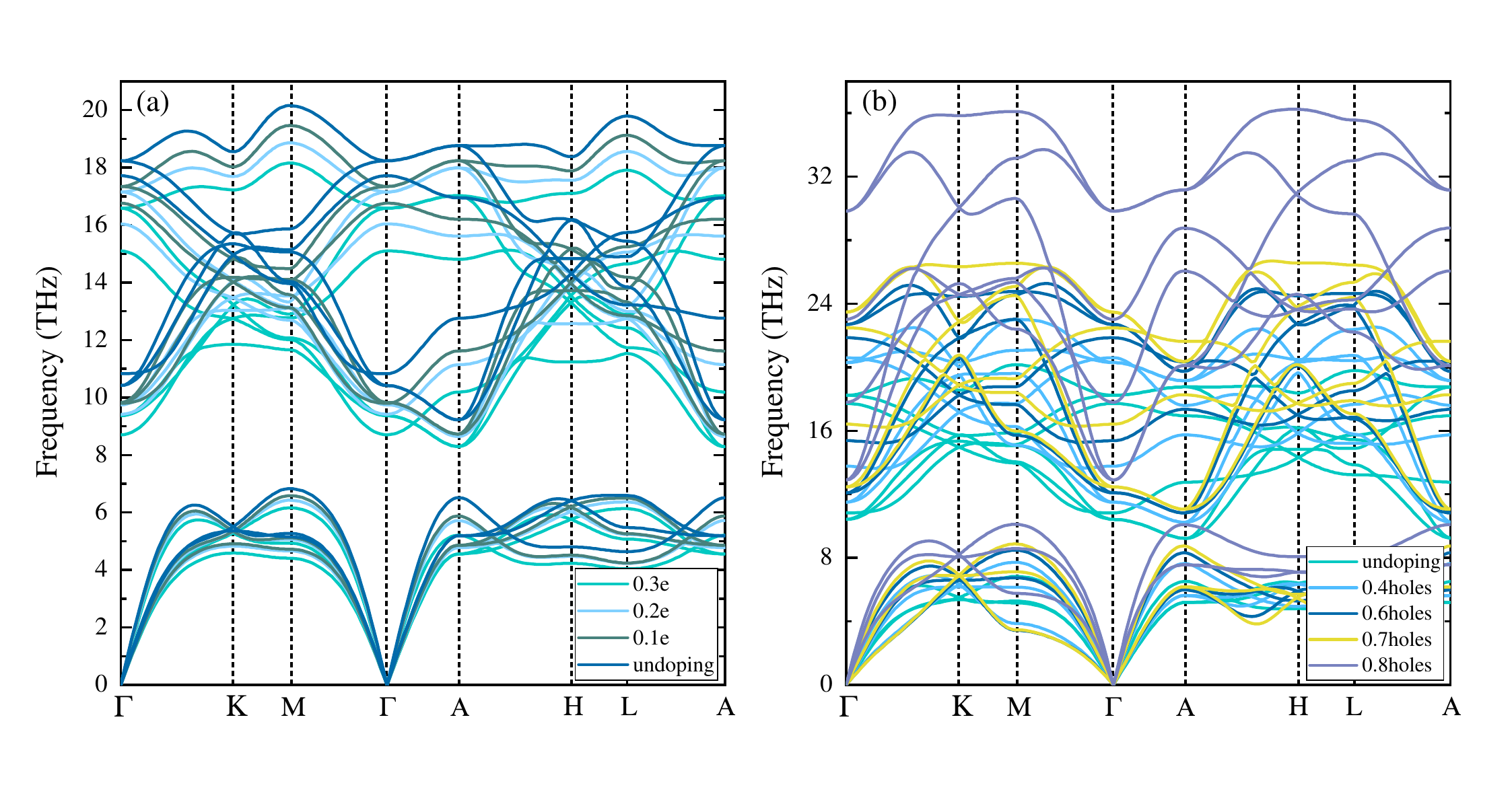}
\caption{
%
The phonon spectrum of YB$_2$ is illustrated under different conditions. (a) The phonon spectrum with varying electron concentrations is shown: light green lines represent the addition of 0.3 electrons, light blue lines indicate the addition of 0.2 electrons, dark green lines correspond to the addition of 0.1 electrons, and dark blue lines represent the spectrum without any added electrons for comparison. (b) The phonon spectrum with varying concentrations of holes is depicted: light blue, dark blue, yellow and purple lines represent the addition of 0.4, 0.6, 0.7 holes and 0.8 respectively, while light green lines correspond to the spectrum without any added holes.
\label{fig6}}
\end{figure*}

\begin{figure*}
\includegraphics[width=2.0\columnwidth]{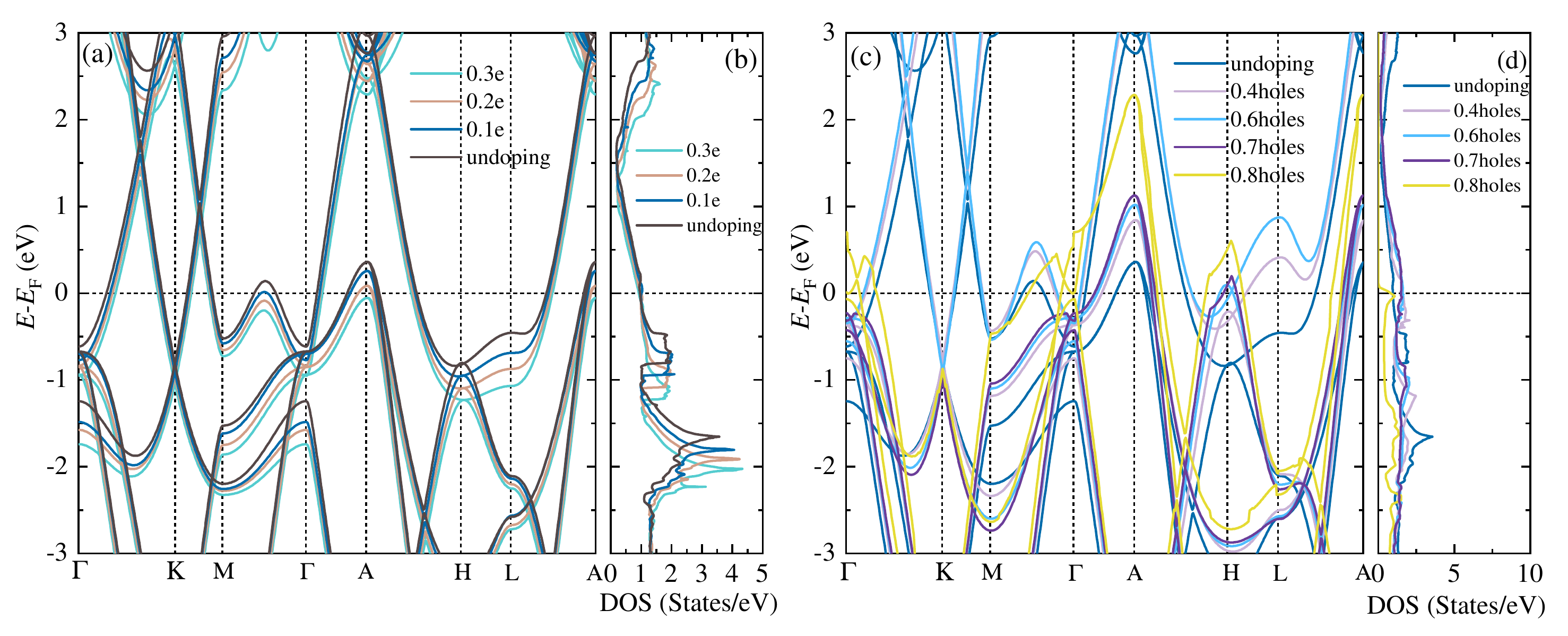}
\caption{
%
The energy bands of YB$_2$ under different concentrations of added electrons and holes are illustrated. (a) The energy band structure with varying electron concentrations: light blue lines represent the addition of 0.3 electrons, pink lines correspond to 0.2 electrons, dark blue lines show the result of adding 0.1 electrons, and dark gray lines serve as a reference without added electrons. (b) The total density of states (DOS) for YB$_2$ under different electron concentrations, with the colored lines corresponding to the same electron concentrations as in (a). (c) The energy band structure with varying concentrations of added holes: light purple, light blue, dark purple and yellow lines represent the addition of 0.4, 0.6, 0.7 and 0.8 holes, respectively, while dark blue lines represent the band structure without added holes. (d) The total DOS of YB$_2$ for varying hole concentrations, where the colored lines correspond to the same conditions as in (c).
\label{fig7}}
\end{figure*}

\begin{figure}
\includegraphics[width=1.0\columnwidth]{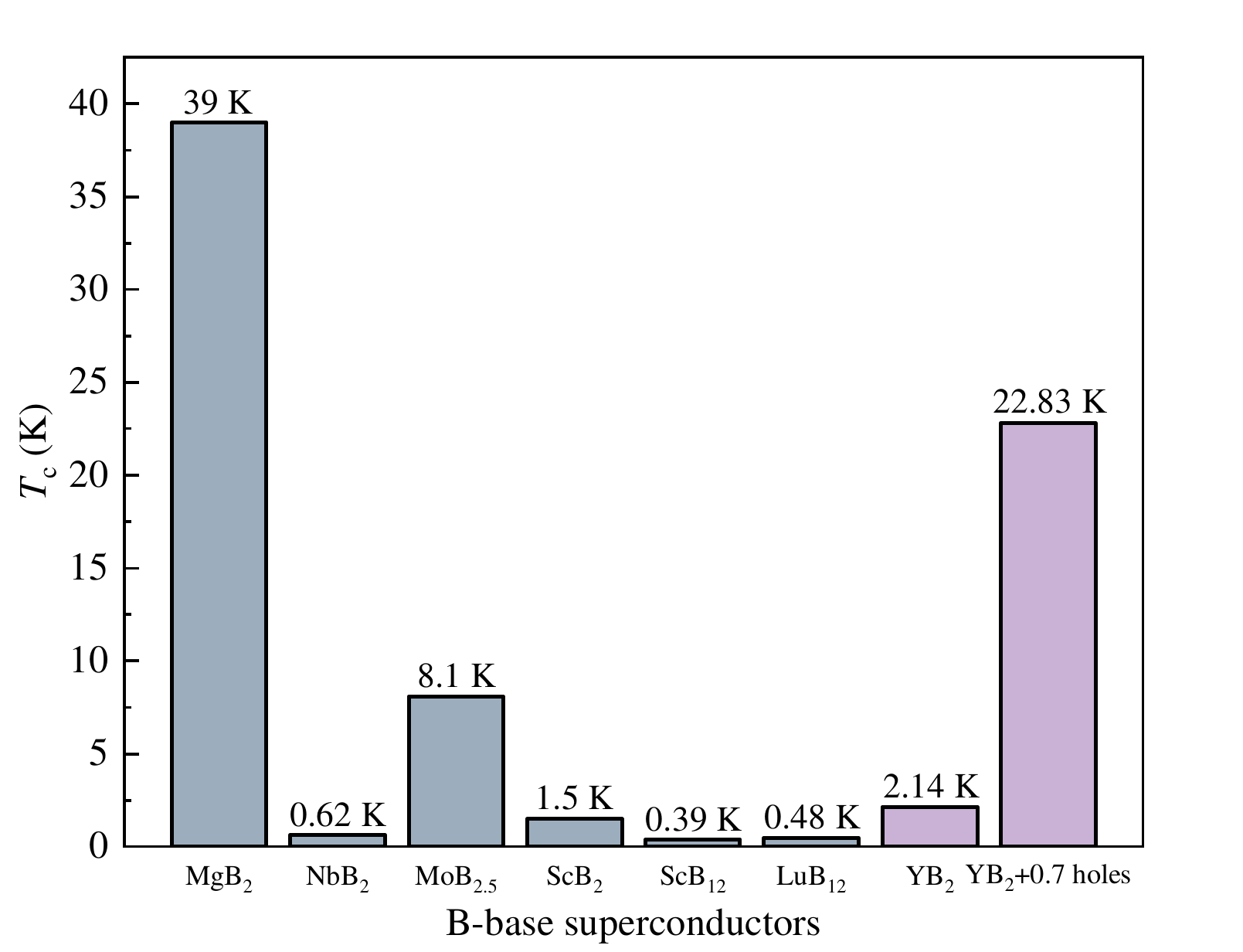}
\caption{
%
The $T_c$ of boron-related compounds at 0 GPa is illustrated, with gray bars representing data from the cited literature: MgB$_2$~\cite{nagamatsu2001superconductivity}; NbB$_2$~\cite{LEYAROVSKA1979249}; MoB$_{2.5}$~\cite{doi:10.1073/pnas.67.1.313}; ScB$_2$~\cite{10.1063/1.4816117}; ScB$_{12}$ and LuB$_{12}$~\cite{doi:10.1126/science.159.3814.530}. The purple bars represent our calculated results.
\label{fig8}}
\end{figure}

\section{III. RESULTS AND DISCUSSION}

YB$_2$, with an AlB$_2$-type structure, belongs to the space group symmetry $P$6/$mmm$ (No.191). Its layered hexagonal crystal structure is illustrated in Fig.~\ref{fig1} (a-b). The overall structure consists of two components: electron-gaining yttrium (Y) atoms and electron-losing boron (B) atoms. The crystal contains two types of sublattices: the upper layer has a hexagonal arrangement, while the other forms a rhombus structure. An intriguing finding is that boron atoms cluster together due to their strong bonding within the same boron layer and with the adjacent yttrium layers above and below. Preliminary analysis, shown in Fig.~\ref{fig1} (c-d), highlights the strong interlayer interactions between boron atoms, which contribute to the superconductivity observed in YB$_2$, as indicated in Fig.~\ref{fig1} (e–f).


The Fermi surface (FS) of YB$_2$ is composed of symmetrically closed ellipsoidal hole sheets at the A symmetry point, along with a uniformly open electron sheet, as depicted in Fig.~\ref{fig2} (a-c). From the FS diagram, the valley degeneracy (N$_v$) is determined to be 2, an exceptionally low value, indicating a low carrier concentration and effective mass in the density of states. The Brillouin zone, including 
the $k$-path, is shown in Fig.~\ref{fig2} (d).

%

The corresponding electronic band structure of YB$_2$ is shown in Fig.~\ref{fig3} (a). As evident from the figure, the orbitals near the Fermi level (E$_F$) are primarily composed of Y-$d$ and B-$p$ orbitals, with minimal contribution from Y-$p$ orbitals. The lower superconducting transition temperature ($T_c$) in YB$_2$ can be attributed to two factors. Firstly, the Van Hove singularity is located 0.79 eV below the Fermi energy, which could contribute to the reduced $T_c$. 
%
%
These factors contribute to a low total density of states (DOS) at E$_F$, around 1.75 states/eV, which helps to explain the low $T_c$ in the following.



The density of states (DOS) of YB$_2$ at 0 GPa is shown in Fig.~\ref{fig3} (b). The DOS is centered in the range of -4 eV to 1 eV. Additionally, the Y-$d$ orbitals overlap with the B-$p$ states and may even overlap with the B-$s$ states, indicating a strong electronic interaction between Y and B atoms. The central region of Fig. S1 highlights the DOS at the Fermi energy level, which consists of B-$p{x,y,z}$ and Y-$d_{xy,x^2,yz}$ orbitals. The amplitude of these components is directly related to the strength of the electron-phonon coupling (EPC). 


Fig.~\ref{fig4} (a) shows the phonon band structure of YB$_2$ at 0 GPa, demonstrating the dynamical stability of the structure. At the $\Gamma$ symmetry point, the slopes of the longitudinal acoustic (LA) and transverse acoustic (TA) modes are 79.88 and 51.90, respectively, corresponding to phonon group velocities of 7.988 km/s for LA and 5.19 km/s for TA. Additionally, a gap of 2.559 THz is observed between the acoustic and optical branches. Consistent with findings for ScB$_{2}$~\cite{10.1063/1.4816117}, these calculations, along with the data in Fig.~\ref{fig2}, suggest a small concentration of holes in YB$_2$ 
which could significantly influence the phonon properties and superconductivity of YB$_2$.


Further examination of the phonon density of states (PDOS) reveals that yttrium (Y) dominates the acoustic branch of the phonon dispersion, while boron (B) is predominant in the optical branch, as shown in Fig.~\ref{fig4} (b), similar to the behavior observed in MgB$_4$~\cite{PhysRevLett.123.077001}. Fig.~\ref{fig4} (c) indicates that high-frequency vibrations in the 12.5 to 20 THz range are the primary contributors to the electron-phonon coupling (EPC) in YB$_2$, with the B element being the main source of these vibrations.



%

As shown in Fig.~\ref{fig5}(a), the McMillan-Allen-Dynes formula and isotropic Eliashberg equations were used to calculate $T_c$. While the results from both methods are comparable, slight variations can be observed.
In addition, experience-based $\mu^*$ ranges from 0.05 to 0.13. Fig.~\ref{fig5}(a) clearly shows a significant decline in $T_c$ from the QE package as $\mu^*$ increases, particularly in the 0.05 to 0.1 range. This indicates that the computed $T_c$ is significantly influenced by the numerical value of $\mu^*$. Therefore, within the computational range, $\mu^*$ was selected to be closest to 0.1, which is the mean value of $T_c$. The different isotropic superconducting gaps $\Delta_0$ of YB$_2$ on the Fermi surface, predicted using isotropic Migdal-Eliashberg equations and depicted as a function of temperature, are shown by the colored lines in Fig.~\ref{fig5}(b). It is evident that the calculated $T_c$ decreases as $\mu^*$ increases. Based on these observations, $\mu^*$ = 0.1 was chosen, leading to a $T_c$ for YB$_2$ of approximately 2.14 K.




As shown in Fig.~\ref{fig2} and Fig.~\ref{fig4}, the analysis revealed that the hole concentration in YB$_2$ is low. To further investigate its superconducting properties, various electron and hole concentrations were computationally inserted, and the resulting $T_c$ values are presented in Fig.~\ref{fig5}(c). On the one hand, with the inclusion of holes, the $T_c$ of YB$_2$ grows monotonically and quickly from 0 to 0.7 to a peak of 22.83 K. But when holes are added, the $T_c$ rapidly drops to even lower than that of pure YB$_2$, with concentrations ranging from 0.7 to 0.8.


On the other hand, the number of electrons added decreases steadily from 0 to 0.3, as indicated by the negative concentrations in Fig.~\ref{fig5}(c). Among the various methods of adding electrons or holes at different concentrations, it is evident that adding holes at a concentration of 0.7 has the most significant effect on increasing the $T_c$ of YB$_2$, making it 7.15 times greater than the $T_c$ without any additions. The steep rise in 
$T_c$ with the addition of holes further supports the analysis that the hole concentration 
in Fig.~\ref{fig2} is insufficient, leading to the lower $T_c$ observed for YB$_2$ in Fig.~\ref{fig5}.  


Moreover, the electron-phonon coupling strength ($\lambda$) and the logarithmically averaged characteristic phonon frequency ($\omega\rm_{log}$), both of which are proportional to $T_c$, are the primary parameters influencing the size of $T_c$, according to Eq.~\ref{eqn1}. The trend of the electron-phonon coupling strength in Fig.~\ref{fig5}(d) for YB$_2$ with varying electron or hole concentrations is similar to that observed in Fig.~\ref{fig5}(c). However, $\lambda$ is smaller than that for pure YB$_2$ after adding either electrons or holes, indicating that $\lambda$ is a key determinant of $T_c$.


The Fig.~\ref{fig6} and Fig.~\ref{fig7} show a comparison of the phonon spectra and energy bands with varying electron and hole concentrations. The phonons gradually become harder as the concentration of holes rises, whereas the addition of electrons completely reverses the situation. Furthermore, with increasing hole concentrations, the region occupied by the peaks of the Eliasberg spectral function at the high-frequency optical phonon and the low-frequency acoustic phonon grows, until both diminish at 0.8 of increased hole concentration. Additionally, it can be demonstrated that the $T_c$ of YB$_2$ grows with increasing hole concentration up to 0.8 because the magnitude of the spectral function is related to the intensity of the electron-phonon coupling. The Fig. S2 and Fig. S3 display these findings. Moreover, the results of the addition of electrons are precisely the reverse of those obtained by the addition of holes.


Surprisingly, the energy bands shown in Fig.~\ref{fig7} (a) and (c) exhibit a similar overall trend. The general downward shift in the energy bands with increasing electron concentration is reversed with the addition of holes. It can be observed that when the concentration of additional electrons grows, the Fermi energy level also increases, whereas the concentration of holes drops. Fig.~\ref{fig7} (b) and (d) illustrate these results, where the total density of states (DOS) at the Fermi energy level decreases gradually with increasing electron concentration, but increases sharply with higher hole concentration. These findings not only confirm the observations from Fig.~\ref{fig5} (c) but also highlight that $T_c$ for YB$_2$ can be further enhanced by increasing the hole concentration.


Another important finding is that this trend is supported by changes in lattice constants shown in Table \ref{tab:t1}. The enhancement of electron-phonon coupling (EPC) induced by pressurization could be related to this result. Lastly, as illustrated in Fig.~\ref{fig8}, the $T_c$ of pure YB$_2$ and YB$_2$ with 0.7 holes added is compared with that of other boron-based compounds. Although the $T_c$ of YB$_2$ is lower than that of MgB$_2$, it is higher than that of all other boron-based superconductors except MoB$_{2.5}$~\cite{doi:10.1073/pnas.67.1.313}. Additionally, with the incorporation of 0.7 holes, the $T_c$ of YB$_2$ surpasses that of MoB$_{2.5}$. This comparison not only demonstrates that pure YB$_2$ has a higher $T_c$ than many boron-related compounds but also highlights that $T_c$ becomes significantly more remarkable when a sufficiently large number of holes are added.

\begin{table}[!ht] 
\renewcommand\arraystretch{1.25} 
\centering 
\caption{
The lattice constant and the corresponding $T_c$ after adding different concentrations of electrons or holes is shown. Negative numbers indicate the addition of electrons, while positive numbers indicate the addition of holes.  
}
\label{tab:t1}
\begin{tabular}{|c|c|c|c|} \hline 

Concentration& a or b (Bohr)& c (Bohr)&  $T_c$ (K) \\ \hline 

-0.3 & 6.43 & 7.59 & 0.82 \\ \hline 

-0.2 & 6.36 & 7.48 & 1.15 \\ \hline  

-0.1 & 6.29 & 7.46 & 2.14 \\ \hline 

0 & 6.22 & 7.29 & 3.19 \\ \hline 

0.1 & 6.15 & 7.21 & 5.26 \\ \hline 

0.2 & 6.07 & 7.14 & 8.16 \\ \hline 

0.3 & 5.99 & 7.06 & 11.03 \\ \hline  

0.4 & 5.90 & 6.99 & 16.41 \\ \hline 

0.5 & 5.80 & 6.92 & 19.70 \\ \hline 

0.6 & 5.71 & 6.85 & 21.37 \\ \hline 

0.7 & 5.62 & 6.77 & 22.83 \\ \hline 

0.8 & 5.53 & 6.71 & 6.22 \\ \hline

\end{tabular}
\end{table}

\section{IV. CONCLUSIONS AND DISCUSSIONS}
%
%
In summary, we determine the precise range of the critical temperature ($T_c$) for YB$_2$ crystal and investigate how electron and hole concentration influence superconducting properties. We also find that varying concentrations of additional holes and electrons can affect $T_c$. Based on the characteristics of bulk YB$_2$, we draw the following conclusions:


(i) When the parameter $\mu^*$ is 0.1, the $T_c$ for YB$_2$ is identified as 2.14 K. Furthermore, $T_c$ is expected to increase as $\mu^*$ decreases.

(ii) Similar to ScB$_2$, the superconducting properties of YB$_2$ are likely due to the close association of atoms in the B layer. The relatively low $T_c$ is primarily attributed to a low concentration of holes 
rather than a lack of electrons.  

(iii) The calculated $T_c$ increases steadily with hole concentrations in the range of 0 to 0.8, reaching 22.83 K at 0.7. 
In contrast, the $T_c$ decreases with added electron concentrations. This offers suggestions for enhancing the $T_c$ of B-based materials that resemble MgB$_2$. Additionally, The $T_c$ value shows an inverse correlation with the lattice constant of the crystal.

\section{V. SUPPLEMENTARY MATERIAL}

The data of the Supplementary Material that support the findings of this study are available at [http://doi.org].

\section{ACKNOWLEDGMENTS}
We acknowledge the support from the National Natural Science Foundation of China 
(No.12104356 and No.52250191). 
Z.G. acknowledges the support of the Fundamental Research Funds for the Central 
Universities. 
The work is supported by the Key Research and Development Program of the Ministry of 
Science and Technology under Grant No.2023YFB4604100.
We also acknowledge the support by HPC Platform, Xi’an Jiaotong University.

{\textbf{AUTHOR DECLARATIONS}

\textbf{Conflict of Interest}

The authors have no conflicts to disclose.

\textbf{DATA AVAILABILITY}

The data that support the findings of this study are available from the corresponding authors upon reasonable request.}

 \bibliography{YB2} 

\begin{thebibliography}{32}%
\makeatletter
\providecommand \@ifxundefined [1]{%
 \@ifx{#1\undefined}
}%
\providecommand \@ifnum [1]{%
 \ifnum #1\expandafter \@firstoftwo
 \else \expandafter \@secondoftwo
 \fi
}%
\providecommand \@ifx [1]{%
 \ifx #1\expandafter \@firstoftwo
 \else \expandafter \@secondoftwo
 \fi
}%
\providecommand \natexlab [1]{#1}%
\providecommand \enquote  [1]{``#1''}%
\providecommand \bibnamefont  [1]{#1}%
\providecommand \bibfnamefont [1]{#1}%
\providecommand \citenamefont [1]{#1}%
\providecommand \href@noop [0]{\@secondoftwo}%
\providecommand \href [0]{\begingroup \@sanitize@url \@href}%
\providecommand \@href[1]{\@@startlink{#1}\@@href}%
\providecommand \@@href[1]{\endgroup#1\@@endlink}%
\providecommand \@sanitize@url [0]{\catcode `\\12\catcode `\$12\catcode `\&12\catcode `\#12\catcode `\^12\catcode `\_12\catcode `\%12\relax}%
\providecommand \@@startlink[1]{}%
\providecommand \@@endlink[0]{}%
\providecommand \url  [0]{\begingroup\@sanitize@url \@url }%
\providecommand \@url [1]{\endgroup\@href {#1}{\urlprefix }}%
\providecommand \urlprefix  [0]{URL }%
\providecommand \Eprint [0]{\href }%
\providecommand \doibase [0]{http://dx.doi.org/}%
\providecommand \selectlanguage [0]{\@gobble}%
\providecommand \bibinfo  [0]{\@secondoftwo}%
\providecommand \bibfield  [0]{\@secondoftwo}%
\providecommand \translation [1]{[#1]}%
\providecommand \BibitemOpen [0]{}%
\providecommand \bibitemStop [0]{}%
\providecommand \bibitemNoStop [0]{.\EOS\space}%
\providecommand \EOS [0]{\spacefactor3000\relax}%
\providecommand \BibitemShut  [1]{\csname bibitem#1\endcsname}%
\let\auto@bib@innerbib\@empty
\bibitem [{\citenamefont {Autler}\ \emph {et~al.}(1962)\citenamefont {Autler}, \citenamefont {Rosenblum},\ and\ \citenamefont {Gooen}}]{PhysRevLett.9.489}%
  \BibitemOpen
  \bibfield  {author} {\bibinfo {author} {\bibfnamefont {S.~H.}\ \bibnamefont {Autler}}, \bibinfo {author} {\bibfnamefont {E.~S.}\ \bibnamefont {Rosenblum}}, \ and\ \bibinfo {author} {\bibfnamefont {K.~H.}\ \bibnamefont {Gooen}},\ }\bibfield  {title} {\enquote {\bibinfo {title} {{High-field superconductivity in Niobium}},}\ }\href {\doibase 10.1103/PhysRevLett.9.489} {\bibfield  {journal} {\bibinfo  {journal} {Phys. Rev. Lett.}\ }\textbf {\bibinfo {volume} {9}},\ \bibinfo {pages} {489--493} (\bibinfo {year} {1962})}\BibitemShut {NoStop}%
\bibitem [{\citenamefont {Chang}\ and\ \citenamefont {Cohen}(1986)}]{PhysRevB.34.4552}%
  \BibitemOpen
  \bibfield  {author} {\bibinfo {author} {\bibfnamefont {K.~J.}\ \bibnamefont {Chang}}\ and\ \bibinfo {author} {\bibfnamefont {Marvin~L.}\ \bibnamefont {Cohen}},\ }\bibfield  {title} {\enquote {\bibinfo {title} {{Electron-phonon interactions and superconductivity in Si, Ge, and Sn}},}\ }\href {\doibase 10.1103/PhysRevB.34.4552} {\bibfield  {journal} {\bibinfo  {journal} {Phys. Rev. B}\ }\textbf {\bibinfo {volume} {34}},\ \bibinfo {pages} {4552--4557} (\bibinfo {year} {1986})}\BibitemShut {NoStop}%
\bibitem [{\citenamefont {Matthias}\ \emph {et~al.}(1968{\natexlab{a}})\citenamefont {Matthias}, \citenamefont {Geballe}, \citenamefont {Andres}, \citenamefont {Corenzwit}, \citenamefont {Hull},\ and\ \citenamefont {Maita}}]{matthias1968superconductivity}%
  \BibitemOpen
  \bibfield  {author} {\bibinfo {author} {\bibfnamefont {BT}~\bibnamefont {Matthias}}, \bibinfo {author} {\bibfnamefont {TH}~\bibnamefont {Geballe}}, \bibinfo {author} {\bibfnamefont {K}~\bibnamefont {Andres}}, \bibinfo {author} {\bibfnamefont {E}~\bibnamefont {Corenzwit}}, \bibinfo {author} {\bibfnamefont {GW}~\bibnamefont {Hull}}, \ and\ \bibinfo {author} {\bibfnamefont {JP}~\bibnamefont {Maita}},\ }\bibfield  {title} {\enquote {\bibinfo {title} {Superconductivity and antiferromagnetism in boron-rich lattices},}\ }\href {\doibase 10.1126/science.159.3814.530} {\bibfield  {journal} {\bibinfo  {journal} {Science}\ }\textbf {\bibinfo {volume} {159}},\ \bibinfo {pages} {530--530} (\bibinfo {year} {1968}{\natexlab{a}})}\BibitemShut {NoStop}%
\bibitem [{\citenamefont {Drozdov}\ \emph {et~al.}(2019)\citenamefont {Drozdov}, \citenamefont {Kong}, \citenamefont {Minkov}, \citenamefont {Besedin}, \citenamefont {Kuzovnikov}, \citenamefont {Mozaffari}, \citenamefont {Balicas}, \citenamefont {Balakirev}, \citenamefont {Graf}, \citenamefont {Prakapenka}, \citenamefont {Greenberg}, \citenamefont {Knyazev}, \citenamefont {Tkacz},\ and\ \citenamefont {Eremets}}]{drozdov2019superconductivity}%
  \BibitemOpen
  \bibfield  {author} {\bibinfo {author} {\bibfnamefont {A.~P.}\ \bibnamefont {Drozdov}}, \bibinfo {author} {\bibfnamefont {P.~P.}\ \bibnamefont {Kong}}, \bibinfo {author} {\bibfnamefont {V.~S.}\ \bibnamefont {Minkov}}, \bibinfo {author} {\bibfnamefont {S.~P.}\ \bibnamefont {Besedin}}, \bibinfo {author} {\bibfnamefont {M.~A.}\ \bibnamefont {Kuzovnikov}}, \bibinfo {author} {\bibfnamefont {S.}~\bibnamefont {Mozaffari}}, \bibinfo {author} {\bibfnamefont {L.}~\bibnamefont {Balicas}}, \bibinfo {author} {\bibfnamefont {F.~F.}\ \bibnamefont {Balakirev}}, \bibinfo {author} {\bibfnamefont {D.~E.}\ \bibnamefont {Graf}}, \bibinfo {author} {\bibfnamefont {V.~B.}\ \bibnamefont {Prakapenka}}, \bibinfo {author} {\bibfnamefont {E.}~\bibnamefont {Greenberg}}, \bibinfo {author} {\bibfnamefont {D.~A.}\ \bibnamefont {Knyazev}}, \bibinfo {author} {\bibfnamefont {M.}~\bibnamefont {Tkacz}}, \ and\ \bibinfo {author} {\bibfnamefont {M.~I.}\ \bibnamefont {Eremets}},\ }\bibfield  {title} {\enquote {\bibinfo {title} {{Superconductivity
  at 250 K in lanthanum hydride under high pressures}},}\ }\href {https://doi.org/10.1038/s41586-019-1201-8} {\bibfield  {journal} {\bibinfo  {journal} {Nature}\ }\textbf {\bibinfo {volume} {569}},\ \bibinfo {pages} {528--531} (\bibinfo {year} {2019})}\BibitemShut {NoStop}%
\bibitem [{\citenamefont {Cross}\ \emph {et~al.}(2024)\citenamefont {Cross}, \citenamefont {Buhot}, \citenamefont {Brooks}, \citenamefont {Thomas}, \citenamefont {Kleppe}, \citenamefont {Lord},\ and\ \citenamefont {Friedemann}}]{cross2024high}%
  \BibitemOpen
  \bibfield  {author} {\bibinfo {author} {\bibfnamefont {Sam}\ \bibnamefont {Cross}}, \bibinfo {author} {\bibfnamefont {Jonathan}\ \bibnamefont {Buhot}}, \bibinfo {author} {\bibfnamefont {Annabelle}\ \bibnamefont {Brooks}}, \bibinfo {author} {\bibfnamefont {William}\ \bibnamefont {Thomas}}, \bibinfo {author} {\bibfnamefont {Annette}\ \bibnamefont {Kleppe}}, \bibinfo {author} {\bibfnamefont {Oliver}\ \bibnamefont {Lord}}, \ and\ \bibinfo {author} {\bibfnamefont {Sven}\ \bibnamefont {Friedemann}},\ }\bibfield  {title} {\enquote {\bibinfo {title} {{High-temperature superconductivity in ${\mathrm{La}}_{4}{\mathrm{H}}_{23}$ below 100 GPa}},}\ }\href {\doibase 10.1103/PhysRevB.109.L020503} {\bibfield  {journal} {\bibinfo  {journal} {Phys. Rev. B}\ }\textbf {\bibinfo {volume} {109}},\ \bibinfo {pages} {L020503} (\bibinfo {year} {2024})}\BibitemShut {NoStop}%
\bibitem [{\citenamefont {Chen}\ \emph {et~al.}(2024)\citenamefont {Chen}, \citenamefont {Guo}, \citenamefont {Wang}, \citenamefont {Wu}, \citenamefont {Chen}, \citenamefont {Huang},\ and\ \citenamefont {Cui}}]{chen2024synthesis}%
  \BibitemOpen
  \bibfield  {author} {\bibinfo {author} {\bibfnamefont {Su}~\bibnamefont {Chen}}, \bibinfo {author} {\bibfnamefont {Jianning}\ \bibnamefont {Guo}}, \bibinfo {author} {\bibfnamefont {Yulong}\ \bibnamefont {Wang}}, \bibinfo {author} {\bibfnamefont {Xinyue}\ \bibnamefont {Wu}}, \bibinfo {author} {\bibfnamefont {Wuhao}\ \bibnamefont {Chen}}, \bibinfo {author} {\bibfnamefont {Xiaoli}\ \bibnamefont {Huang}}, \ and\ \bibinfo {author} {\bibfnamefont {Tian}\ \bibnamefont {Cui}},\ }\bibfield  {title} {\enquote {\bibinfo {title} {{Synthesis and superconductivity in $(\mathrm{La},\mathrm{Ca}){\mathrm{H}}_{10}$ under high pressure}},}\ }\href {\doibase 10.1103/PhysRevB.109.224510} {\bibfield  {journal} {\bibinfo  {journal} {Phys. Rev. B}\ }\textbf {\bibinfo {volume} {109}},\ \bibinfo {pages} {224510} (\bibinfo {year} {2024})}\BibitemShut {NoStop}%
\bibitem [{\citenamefont {Nagamatsu}\ \emph {et~al.}(2001)\citenamefont {Nagamatsu}, \citenamefont {Nakagawa}, \citenamefont {Muranaka}, \citenamefont {Zenitani},\ and\ \citenamefont {Akimitsu}}]{nagamatsu2001superconductivity}%
  \BibitemOpen
  \bibfield  {author} {\bibinfo {author} {\bibfnamefont {Jun}\ \bibnamefont {Nagamatsu}}, \bibinfo {author} {\bibfnamefont {Norimasa}\ \bibnamefont {Nakagawa}}, \bibinfo {author} {\bibfnamefont {Takahiro}\ \bibnamefont {Muranaka}}, \bibinfo {author} {\bibfnamefont {Yuji}\ \bibnamefont {Zenitani}}, \ and\ \bibinfo {author} {\bibfnamefont {Jun}\ \bibnamefont {Akimitsu}},\ }\bibfield  {title} {\enquote {\bibinfo {title} {{Superconductivity at 39 K in magnesium diboride}},}\ }\href {https://doi.org/10.1038/35065039} {\bibfield  {journal} {\bibinfo  {journal} {Nature}\ }\textbf {\bibinfo {volume} {410}},\ \bibinfo {pages} {63--64} (\bibinfo {year} {2001})}\BibitemShut {NoStop}%
\bibitem [{\citenamefont {Leyarovska}\ and\ \citenamefont {Leyarovski}(1979)}]{LEYAROVSKA1979249}%
  \BibitemOpen
  \bibfield  {author} {\bibinfo {author} {\bibfnamefont {L}~\bibnamefont {Leyarovska}}\ and\ \bibinfo {author} {\bibfnamefont {E}~\bibnamefont {Leyarovski}},\ }\bibfield  {title} {\enquote {\bibinfo {title} {{A search for superconductivity below 1 K in transition metal borides}},}\ }\href {\doibase https://doi.org/10.1016/0022-5088(79)90100-0} {\bibfield  {journal} {\bibinfo  {journal} {J. Less-Common Met.}\ }\textbf {\bibinfo {volume} {67}},\ \bibinfo {pages} {249--255} (\bibinfo {year} {1979})}\BibitemShut {NoStop}%
\bibitem [{\citenamefont {Sichkar}\ and\ \citenamefont {Antonov}(2013)}]{10.1063/1.4816117}%
  \BibitemOpen
  \bibfield  {author} {\bibinfo {author} {\bibfnamefont {S.~M.}\ \bibnamefont {Sichkar}}\ and\ \bibinfo {author} {\bibfnamefont {V.~N.}\ \bibnamefont {Antonov}},\ }\bibfield  {title} {\enquote {\bibinfo {title} {{Electronic structure, phonon spectra and electron–phonon interaction in ScB$_2$}},}\ }\href {\doibase 10.1063/1.4816117} {\bibfield  {journal} {\bibinfo  {journal} {Low Temp. Phys.}\ }\textbf {\bibinfo {volume} {39}},\ \bibinfo {pages} {595--601} (\bibinfo {year} {2013})}\BibitemShut {NoStop}%
\bibitem [{\citenamefont {Cooper}\ \emph {et~al.}(1970)\citenamefont {Cooper}, \citenamefont {Corenzwit}, \citenamefont {Longinotti}, \citenamefont {Matthias},\ and\ \citenamefont {Zachariasen}}]{doi:10.1073/pnas.67.1.313}%
  \BibitemOpen
  \bibfield  {author} {\bibinfo {author} {\bibfnamefont {A.~S.}\ \bibnamefont {Cooper}}, \bibinfo {author} {\bibfnamefont {E.}~\bibnamefont {Corenzwit}}, \bibinfo {author} {\bibfnamefont {L.~D.}\ \bibnamefont {Longinotti}}, \bibinfo {author} {\bibfnamefont {B.~T.}\ \bibnamefont {Matthias}}, \ and\ \bibinfo {author} {\bibfnamefont {W.~H.}\ \bibnamefont {Zachariasen}},\ }\bibfield  {title} {\enquote {\bibinfo {title} {Superconductivity: The transition temperature peak below four electrons per atom},}\ }\href {\doibase 10.1073/pnas.67.1.313} {\bibfield  {journal} {\bibinfo  {journal} {Proc. Natl. Acad. Sci.}\ }\textbf {\bibinfo {volume} {67}},\ \bibinfo {pages} {313--319} (\bibinfo {year} {1970})}\BibitemShut {NoStop}%
\bibitem [{\citenamefont {Liu}\ \emph {et~al.}(2022)\citenamefont {Liu}, \citenamefont {Huang}, \citenamefont {Song}, \citenamefont {Wang}, \citenamefont {Zhang}, \citenamefont {Lv}, \citenamefont {Liu}, \citenamefont {Zhang}, \citenamefont {Cho},\ and\ \citenamefont {Jia}}]{PhysRevB.106.064507}%
  \BibitemOpen
  \bibfield  {author} {\bibinfo {author} {\bibfnamefont {Xiaohan}\ \bibnamefont {Liu}}, \bibinfo {author} {\bibfnamefont {Xiaowei}\ \bibnamefont {Huang}}, \bibinfo {author} {\bibfnamefont {Peng}\ \bibnamefont {Song}}, \bibinfo {author} {\bibfnamefont {Chongze}\ \bibnamefont {Wang}}, \bibinfo {author} {\bibfnamefont {Liying}\ \bibnamefont {Zhang}}, \bibinfo {author} {\bibfnamefont {Peng}\ \bibnamefont {Lv}}, \bibinfo {author} {\bibfnamefont {Liangliang}\ \bibnamefont {Liu}}, \bibinfo {author} {\bibfnamefont {Weifeng}\ \bibnamefont {Zhang}}, \bibinfo {author} {\bibfnamefont {Jun-Hyung}\ \bibnamefont {Cho}}, \ and\ \bibinfo {author} {\bibfnamefont {Yu}~\bibnamefont {Jia}},\ }\bibfield  {title} {\enquote {\bibinfo {title} {{Strong electron-phonon coupling superconductivity in compressed $\ensuremath{\alpha}\text{\ensuremath{-}}{\mathrm{MoB}}_{2}$ induced by double Van Hove singularities}},}\ }\href {\doibase 10.1103/PhysRevB.106.064507} {\bibfield  {journal} {\bibinfo  {journal} {Phys. Rev. B}\ }\textbf {\bibinfo
  {volume} {106}},\ \bibinfo {pages} {064507} (\bibinfo {year} {2022})}\BibitemShut {NoStop}%
\bibitem [{\citenamefont {Choi}\ \emph {et~al.}(2009)\citenamefont {Choi}, \citenamefont {Louie},\ and\ \citenamefont {Cohen}}]{PhysRevB.80.064503}%
  \BibitemOpen
  \bibfield  {author} {\bibinfo {author} {\bibfnamefont {Hyoung~Joon}\ \bibnamefont {Choi}}, \bibinfo {author} {\bibfnamefont {Steven~G.}\ \bibnamefont {Louie}}, \ and\ \bibinfo {author} {\bibfnamefont {Marvin~L.}\ \bibnamefont {Cohen}},\ }\bibfield  {title} {\enquote {\bibinfo {title} {{Prediction of superconducting properties of ${\text{CaB}}_{2}$ using anisotropic Eliashberg theory}},}\ }\href {\doibase 10.1103/PhysRevB.80.064503} {\bibfield  {journal} {\bibinfo  {journal} {Phys. Rev. B}\ }\textbf {\bibinfo {volume} {80}},\ \bibinfo {pages} {064503} (\bibinfo {year} {2009})}\BibitemShut {NoStop}%
\bibitem [{\citenamefont {Medvedeva}\ \emph {et~al.}(2001)\citenamefont {Medvedeva}, \citenamefont {Ivanovskii}, \citenamefont {Medvedeva},\ and\ \citenamefont {Freeman}}]{PhysRevB.64.020502}%
  \BibitemOpen
  \bibfield  {author} {\bibinfo {author} {\bibfnamefont {N.~I.}\ \bibnamefont {Medvedeva}}, \bibinfo {author} {\bibfnamefont {A.~L.}\ \bibnamefont {Ivanovskii}}, \bibinfo {author} {\bibfnamefont {J.~E.}\ \bibnamefont {Medvedeva}}, \ and\ \bibinfo {author} {\bibfnamefont {A.~J.}\ \bibnamefont {Freeman}},\ }\bibfield  {title} {\enquote {\bibinfo {title} {{Electronic structure of superconducting ${\mathrm{MgB}}_{2}$ and related binary and ternary borides}},}\ }\href {\doibase 10.1103/PhysRevB.64.020502} {\bibfield  {journal} {\bibinfo  {journal} {Phys. Rev. B}\ }\textbf {\bibinfo {volume} {64}},\ \bibinfo {pages} {020502} (\bibinfo {year} {2001})}\BibitemShut {NoStop}%
\bibitem [{\citenamefont {Chen}\ \emph {et~al.}(2001)\citenamefont {Chen}, \citenamefont {Tu}, \citenamefont {He}, \citenamefont {Dai},\ and\ \citenamefont {Wu}}]{XLChen_2001}%
  \BibitemOpen
  \bibfield  {author} {\bibinfo {author} {\bibfnamefont {X~L}\ \bibnamefont {Chen}}, \bibinfo {author} {\bibfnamefont {Q~Y}\ \bibnamefont {Tu}}, \bibinfo {author} {\bibfnamefont {M}~\bibnamefont {He}}, \bibinfo {author} {\bibfnamefont {L}~\bibnamefont {Dai}}, \ and\ \bibinfo {author} {\bibfnamefont {L}~\bibnamefont {Wu}},\ }\bibfield  {title} {\enquote {\bibinfo {title} {{The bond ionicity of MB$_2$ (M = Mg, Ti, V, Cr, Mn, Zr, Hf, Ta, Al and Y)}},}\ }\href {\doibase 10.1088/0953-8984/13/29/105} {\bibfield  {journal} {\bibinfo  {journal} {J. Phys.: Condens. Matter}\ }\textbf {\bibinfo {volume} {13}},\ \bibinfo {pages} {L723} (\bibinfo {year} {2001})}\BibitemShut {NoStop}%
\bibitem [{\citenamefont {Li}\ \emph {et~al.}(2024)\citenamefont {Li}, \citenamefont {Xu}, \citenamefont {Li}, \citenamefont {Li}, \citenamefont {Li}, \citenamefont {Watanabe}, \citenamefont {Taniguchi}, \citenamefont {Tong}, \citenamefont {Shen}, \citenamefont {Lu}, \citenamefont {Jia}, \citenamefont {Wu}, \citenamefont {Liu},\ and\ \citenamefont {Li}}]{li2024tunable}%
  \BibitemOpen
  \bibfield  {author} {\bibinfo {author} {\bibfnamefont {Chushan}\ \bibnamefont {Li}}, \bibinfo {author} {\bibfnamefont {Fan}\ \bibnamefont {Xu}}, \bibinfo {author} {\bibfnamefont {Bohao}\ \bibnamefont {Li}}, \bibinfo {author} {\bibfnamefont {Jiayi}\ \bibnamefont {Li}}, \bibinfo {author} {\bibfnamefont {Guoan}\ \bibnamefont {Li}}, \bibinfo {author} {\bibfnamefont {Kenji}\ \bibnamefont {Watanabe}}, \bibinfo {author} {\bibfnamefont {Takashi}\ \bibnamefont {Taniguchi}}, \bibinfo {author} {\bibfnamefont {Bingbing}\ \bibnamefont {Tong}}, \bibinfo {author} {\bibfnamefont {Jie}\ \bibnamefont {Shen}}, \bibinfo {author} {\bibfnamefont {Li}~\bibnamefont {Lu}}, \bibinfo {author} {\bibfnamefont {Jinfeng}\ \bibnamefont {Jia}}, \bibinfo {author} {\bibfnamefont {Fengcheng}\ \bibnamefont {Wu}}, \bibinfo {author} {\bibfnamefont {Xiaoxue}\ \bibnamefont {Liu}}, \ and\ \bibinfo {author} {\bibfnamefont {Tingxin}\ \bibnamefont {Li}},\ }\bibfield  {title} {\enquote {\bibinfo {title} {Tunable superconductivity in electron-and
  hole-doped bernal bilayer graphene},}\ }\href {\doibase 10.1038/s41586-024-07584-w} {\bibfield  {journal} {\bibinfo  {journal} {Nature}\ ,\ \bibinfo {pages} {1--7}} (\bibinfo {year} {2024})}\BibitemShut {NoStop}%
\bibitem [{\citenamefont {Rudenko}\ \emph {et~al.}(2024)\citenamefont {Rudenko}, \citenamefont {Badrtdinov}, \citenamefont {Abrikosov},\ and\ \citenamefont {Katsnelson}}]{rudenko2024strong}%
  \BibitemOpen
  \bibfield  {author} {\bibinfo {author} {\bibfnamefont {Alexander~N.}\ \bibnamefont {Rudenko}}, \bibinfo {author} {\bibfnamefont {Danis~I.}\ \bibnamefont {Badrtdinov}}, \bibinfo {author} {\bibfnamefont {Igor~A.}\ \bibnamefont {Abrikosov}}, \ and\ \bibinfo {author} {\bibfnamefont {Mikhail~I.}\ \bibnamefont {Katsnelson}},\ }\bibfield  {title} {\enquote {\bibinfo {title} {{Strong electron-phonon coupling and phonon-induced superconductivity in tetragonal ${\mathrm{C}}_{3}{\mathrm{N}}_{4}$ with hole doping}},}\ }\href {\doibase 10.1103/PhysRevB.109.014502} {\bibfield  {journal} {\bibinfo  {journal} {Phys. Rev. B}\ }\textbf {\bibinfo {volume} {109}},\ \bibinfo {pages} {014502} (\bibinfo {year} {2024})}\BibitemShut {NoStop}%
\bibitem [{\citenamefont {Choi}\ \emph {et~al.}(2024)\citenamefont {Choi}, \citenamefont {Tu}, \citenamefont {Nag}, \citenamefont {Tam}, \citenamefont {Tippireddy}, \citenamefont {Agrestini}, \citenamefont {Lin}, \citenamefont {Garcia-Fernandez}, \citenamefont {Jin},\ and\ \citenamefont {Zhou}}]{choi2024unified}%
  \BibitemOpen
  \bibfield  {author} {\bibinfo {author} {\bibfnamefont {Jaewon}\ \bibnamefont {Choi}}, \bibinfo {author} {\bibfnamefont {Sijia}\ \bibnamefont {Tu}}, \bibinfo {author} {\bibfnamefont {Abhishek}\ \bibnamefont {Nag}}, \bibinfo {author} {\bibfnamefont {Charles~C}\ \bibnamefont {Tam}}, \bibinfo {author} {\bibfnamefont {Sahil}\ \bibnamefont {Tippireddy}}, \bibinfo {author} {\bibfnamefont {Stefano}\ \bibnamefont {Agrestini}}, \bibinfo {author} {\bibfnamefont {Zefeng}\ \bibnamefont {Lin}}, \bibinfo {author} {\bibfnamefont {Mirian}\ \bibnamefont {Garcia-Fernandez}}, \bibinfo {author} {\bibfnamefont {Kui}\ \bibnamefont {Jin}}, \ and\ \bibinfo {author} {\bibfnamefont {Ke-Jin}\ \bibnamefont {Zhou}},\ }\bibfield  {title} {\enquote {\bibinfo {title} {Unified description of charge density waves in electron-and hole-doped cuprate superconductors},}\ }\href@noop {} {\bibfield  {journal} {\bibinfo  {journal} {arXiv preprint arXiv:2407.15750}\ } (\bibinfo {year} {2024})}\BibitemShut {NoStop}%
\bibitem [{\citenamefont {Flores-Livas}\ \emph {et~al.}(2020)\citenamefont {Flores-Livas}, \citenamefont {Boeri}, \citenamefont {Sanna}, \citenamefont {Profeta}, \citenamefont {Arita},\ and\ \citenamefont {Eremets}}]{FLORESLIVAS20201}%
  \BibitemOpen
  \bibfield  {author} {\bibinfo {author} {\bibfnamefont {José~A.}\ \bibnamefont {Flores-Livas}}, \bibinfo {author} {\bibfnamefont {Lilia}\ \bibnamefont {Boeri}}, \bibinfo {author} {\bibfnamefont {Antonio}\ \bibnamefont {Sanna}}, \bibinfo {author} {\bibfnamefont {Gianni}\ \bibnamefont {Profeta}}, \bibinfo {author} {\bibfnamefont {Ryotaro}\ \bibnamefont {Arita}}, \ and\ \bibinfo {author} {\bibfnamefont {Mikhail}\ \bibnamefont {Eremets}},\ }\bibfield  {title} {\enquote {\bibinfo {title} {A perspective on conventional high-temperature superconductors at high pressure: Methods and materials},}\ }\href {\doibase https://doi.org/10.1016/j.physrep.2020.02.003} {\bibfield  {journal} {\bibinfo  {journal} {Physics Reports}\ }\textbf {\bibinfo {volume} {856}},\ \bibinfo {pages} {1--78} (\bibinfo {year} {2020})}\BibitemShut {NoStop}%
\bibitem [{\citenamefont {Hilleke}\ and\ \citenamefont {Zurek}(2022)}]{10.1063/5.0077748}%
  \BibitemOpen
  \bibfield  {author} {\bibinfo {author} {\bibfnamefont {Katerina~P.}\ \bibnamefont {Hilleke}}\ and\ \bibinfo {author} {\bibfnamefont {Eva}\ \bibnamefont {Zurek}},\ }\bibfield  {title} {\enquote {\bibinfo {title} {{Tuning chemical precompression: Theoretical design and crystal chemistry of novel hydrides in the quest for warm and light superconductivity at ambient pressures}},}\ }\href {\doibase 10.1063/5.0077748} {\bibfield  {journal} {\bibinfo  {journal} {Journal of Applied Physics}\ }\textbf {\bibinfo {volume} {131}},\ \bibinfo {pages} {070901} (\bibinfo {year} {2022})}\BibitemShut {NoStop}%
\bibitem [{\citenamefont {Allen}\ and\ \citenamefont {Dynes}(1975)}]{PhysRevB.12.905}%
  \BibitemOpen
  \bibfield  {author} {\bibinfo {author} {\bibfnamefont {P.~B.}\ \bibnamefont {Allen}}\ and\ \bibinfo {author} {\bibfnamefont {R.~C.}\ \bibnamefont {Dynes}},\ }\bibfield  {title} {\enquote {\bibinfo {title} {Transition temperature of strong-coupled superconductors reanalyzed},}\ }\href {\doibase 10.1103/PhysRevB.12.905} {\bibfield  {journal} {\bibinfo  {journal} {Phys. Rev. B}\ }\textbf {\bibinfo {volume} {12}},\ \bibinfo {pages} {905--922} (\bibinfo {year} {1975})}\BibitemShut {NoStop}%
\bibitem [{\citenamefont {Kresse}\ and\ \citenamefont {Furthmüller}(1996)}]{KRESSE199615}%
  \BibitemOpen
  \bibfield  {author} {\bibinfo {author} {\bibfnamefont {G.}~\bibnamefont {Kresse}}\ and\ \bibinfo {author} {\bibfnamefont {J.}~\bibnamefont {Furthmüller}},\ }\bibfield  {title} {\enquote {\bibinfo {title} {Efficiency of ab-initio total energy calculations for metals and semiconductors using a plane-wave basis set},}\ }\href {\doibase https://doi.org/10.1016/0927-0256(96)00008-0} {\bibfield  {journal} {\bibinfo  {journal} {Comp Mater Sci}\ }\textbf {\bibinfo {volume} {6}},\ \bibinfo {pages} {15--50} (\bibinfo {year} {1996})}\BibitemShut {NoStop}%
\bibitem [{\citenamefont {Kresse}\ and\ \citenamefont {Furthm\"uller}(1996)}]{PhysRevB.54.11169}%
  \BibitemOpen
  \bibfield  {author} {\bibinfo {author} {\bibfnamefont {G.}~\bibnamefont {Kresse}}\ and\ \bibinfo {author} {\bibfnamefont {J.}~\bibnamefont {Furthm\"uller}},\ }\bibfield  {title} {\enquote {\bibinfo {title} {Efficient iterative schemes for ab initio total-energy calculations using a plane-wave basis set},}\ }\href {\doibase 10.1103/PhysRevB.54.11169} {\bibfield  {journal} {\bibinfo  {journal} {Phys. Rev. B}\ }\textbf {\bibinfo {volume} {54}},\ \bibinfo {pages} {11169--11186} (\bibinfo {year} {1996})}\BibitemShut {NoStop}%
\bibitem [{\citenamefont {Perdew}\ \emph {et~al.}(1996)\citenamefont {Perdew}, \citenamefont {Burke},\ and\ \citenamefont {Ernzerhof}}]{PhysRevLett.77.3865}%
  \BibitemOpen
  \bibfield  {author} {\bibinfo {author} {\bibfnamefont {John~P.}\ \bibnamefont {Perdew}}, \bibinfo {author} {\bibfnamefont {Kieron}\ \bibnamefont {Burke}}, \ and\ \bibinfo {author} {\bibfnamefont {Matthias}\ \bibnamefont {Ernzerhof}},\ }\bibfield  {title} {\enquote {\bibinfo {title} {Generalized gradient approximation made simple},}\ }\href {\doibase 10.1103/PhysRevLett.77.3865} {\bibfield  {journal} {\bibinfo  {journal} {Phys. Rev. Lett.}\ }\textbf {\bibinfo {volume} {77}},\ \bibinfo {pages} {3865--3868} (\bibinfo {year} {1996})}\BibitemShut {NoStop}%
\bibitem [{\citenamefont {White}\ and\ \citenamefont {Bird}(1994)}]{PhysRevB.50.4954}%
  \BibitemOpen
  \bibfield  {author} {\bibinfo {author} {\bibfnamefont {J.~A.}\ \bibnamefont {White}}\ and\ \bibinfo {author} {\bibfnamefont {D.~M.}\ \bibnamefont {Bird}},\ }\bibfield  {title} {\enquote {\bibinfo {title} {Implementation of gradient-corrected exchange-correlation potentials in car-parrinello total-energy calculations},}\ }\href {\doibase 10.1103/PhysRevB.50.4954} {\bibfield  {journal} {\bibinfo  {journal} {Phys. Rev. B}\ }\textbf {\bibinfo {volume} {50}},\ \bibinfo {pages} {4954--4957} (\bibinfo {year} {1994})}\BibitemShut {NoStop}%
\bibitem [{\citenamefont {Wu}\ and\ \citenamefont {Cohen}(2006)}]{PhysRevB.73.235116}%
  \BibitemOpen
  \bibfield  {author} {\bibinfo {author} {\bibfnamefont {Zhigang}\ \bibnamefont {Wu}}\ and\ \bibinfo {author} {\bibfnamefont {R.~E.}\ \bibnamefont {Cohen}},\ }\bibfield  {title} {\enquote {\bibinfo {title} {More accurate generalized gradient approximation for solids},}\ }\href {\doibase 10.1103/PhysRevB.73.235116} {\bibfield  {journal} {\bibinfo  {journal} {Phys. Rev. B}\ }\textbf {\bibinfo {volume} {73}},\ \bibinfo {pages} {235116} (\bibinfo {year} {2006})}\BibitemShut {NoStop}%
\bibitem [{\citenamefont {Kresse}\ and\ \citenamefont {Hafner}(1993)}]{PhysRevB.48.13115}%
  \BibitemOpen
  \bibfield  {author} {\bibinfo {author} {\bibfnamefont {G.}~\bibnamefont {Kresse}}\ and\ \bibinfo {author} {\bibfnamefont {J.}~\bibnamefont {Hafner}},\ }\bibfield  {title} {\enquote {\bibinfo {title} {Ab initio molecular dynamics for open-shell transition metals},}\ }\href {\doibase 10.1103/PhysRevB.48.13115} {\bibfield  {journal} {\bibinfo  {journal} {Phys. Rev. B}\ }\textbf {\bibinfo {volume} {48}},\ \bibinfo {pages} {13115--13118} (\bibinfo {year} {1993})}\BibitemShut {NoStop}%
\bibitem [{\citenamefont {Bl\"ochl}(1994)}]{PhysRevB.50.17953}%
  \BibitemOpen
  \bibfield  {author} {\bibinfo {author} {\bibfnamefont {P.~E.}\ \bibnamefont {Bl\"ochl}},\ }\bibfield  {title} {\enquote {\bibinfo {title} {Projector augmented-wave method},}\ }\href {\doibase 10.1103/PhysRevB.50.17953} {\bibfield  {journal} {\bibinfo  {journal} {Phys. Rev. B}\ }\textbf {\bibinfo {volume} {50}},\ \bibinfo {pages} {17953--17979} (\bibinfo {year} {1994})}\BibitemShut {NoStop}%
\bibitem [{\citenamefont {Giannozzi}\ \emph {et~al.}(2009)\citenamefont {Giannozzi}, \citenamefont {Baroni}, \citenamefont {Bonini}, \citenamefont {Calandra}, \citenamefont {Car}, \citenamefont {Cavazzoni}, \citenamefont {Ceresoli}, \citenamefont {Chiarotti}, \citenamefont {Cococcioni}, \citenamefont {Dabo}, \citenamefont {Corso}, \citenamefont {de~Gironcoli}, \citenamefont {Fabris}, \citenamefont {Fratesi}, \citenamefont {Gebauer}, \citenamefont {Gerstmann}, \citenamefont {Gougoussis}, \citenamefont {Kokalj}, \citenamefont {Lazzeri}, \citenamefont {Martin-Samos}, \citenamefont {Marzari}, \citenamefont {Mauri}, \citenamefont {Mazzarello}, \citenamefont {Paolini}, \citenamefont {Pasquarello}, \citenamefont {Paulatto}, \citenamefont {Sbraccia}, \citenamefont {Scandolo}, \citenamefont {Sclauzero}, \citenamefont {Seitsonen}, \citenamefont {Smogunov}, \citenamefont {Umari},\ and\ \citenamefont {Wentzcovitch}}]{Giannozzi_2009}%
  \BibitemOpen
  \bibfield  {author} {\bibinfo {author} {\bibfnamefont {Paolo}\ \bibnamefont {Giannozzi}}, \bibinfo {author} {\bibfnamefont {Stefano}\ \bibnamefont {Baroni}}, \bibinfo {author} {\bibfnamefont {Nicola}\ \bibnamefont {Bonini}}, \bibinfo {author} {\bibfnamefont {Matteo}\ \bibnamefont {Calandra}}, \bibinfo {author} {\bibfnamefont {Roberto}\ \bibnamefont {Car}}, \bibinfo {author} {\bibfnamefont {Carlo}\ \bibnamefont {Cavazzoni}}, \bibinfo {author} {\bibfnamefont {Davide}\ \bibnamefont {Ceresoli}}, \bibinfo {author} {\bibfnamefont {Guido~L}\ \bibnamefont {Chiarotti}}, \bibinfo {author} {\bibfnamefont {Matteo}\ \bibnamefont {Cococcioni}}, \bibinfo {author} {\bibfnamefont {Ismaila}\ \bibnamefont {Dabo}}, \bibinfo {author} {\bibfnamefont {Andrea~Dal}\ \bibnamefont {Corso}}, \bibinfo {author} {\bibfnamefont {Stefano}\ \bibnamefont {de~Gironcoli}}, \bibinfo {author} {\bibfnamefont {Stefano}\ \bibnamefont {Fabris}}, \bibinfo {author} {\bibfnamefont {Guido}\ \bibnamefont {Fratesi}}, \bibinfo {author} {\bibfnamefont
  {Ralph}\ \bibnamefont {Gebauer}}, \bibinfo {author} {\bibfnamefont {Uwe}\ \bibnamefont {Gerstmann}}, \bibinfo {author} {\bibfnamefont {Christos}\ \bibnamefont {Gougoussis}}, \bibinfo {author} {\bibfnamefont {Anton}\ \bibnamefont {Kokalj}}, \bibinfo {author} {\bibfnamefont {Michele}\ \bibnamefont {Lazzeri}}, \bibinfo {author} {\bibfnamefont {Layla}\ \bibnamefont {Martin-Samos}}, \bibinfo {author} {\bibfnamefont {Nicola}\ \bibnamefont {Marzari}}, \bibinfo {author} {\bibfnamefont {Francesco}\ \bibnamefont {Mauri}}, \bibinfo {author} {\bibfnamefont {Riccardo}\ \bibnamefont {Mazzarello}}, \bibinfo {author} {\bibfnamefont {Stefano}\ \bibnamefont {Paolini}}, \bibinfo {author} {\bibfnamefont {Alfredo}\ \bibnamefont {Pasquarello}}, \bibinfo {author} {\bibfnamefont {Lorenzo}\ \bibnamefont {Paulatto}}, \bibinfo {author} {\bibfnamefont {Carlo}\ \bibnamefont {Sbraccia}}, \bibinfo {author} {\bibfnamefont {Sandro}\ \bibnamefont {Scandolo}}, \bibinfo {author} {\bibfnamefont {Gabriele}\ \bibnamefont {Sclauzero}}, \bibinfo
  {author} {\bibfnamefont {Ari~P}\ \bibnamefont {Seitsonen}}, \bibinfo {author} {\bibfnamefont {Alexander}\ \bibnamefont {Smogunov}}, \bibinfo {author} {\bibfnamefont {Paolo}\ \bibnamefont {Umari}}, \ and\ \bibinfo {author} {\bibfnamefont {Renata~M}\ \bibnamefont {Wentzcovitch}},\ }\bibfield  {title} {\enquote {\bibinfo {title} {Quantum espresso: a modular and open-source software project for quantum simulations of materials},}\ }\href {\doibase 10.1088/0953-8984/21/39/395502} {\bibfield  {journal} {\bibinfo  {journal} {J. Phys.: Condens. Matter}\ }\textbf {\bibinfo {volume} {21}},\ \bibinfo {pages} {395502} (\bibinfo {year} {2009})}\BibitemShut {NoStop}%
\bibitem [{\citenamefont {Schlipf}\ and\ \citenamefont {Gygi}(2015)}]{SCHLIPF201536}%
  \BibitemOpen
  \bibfield  {author} {\bibinfo {author} {\bibfnamefont {Martin}\ \bibnamefont {Schlipf}}\ and\ \bibinfo {author} {\bibfnamefont {François}\ \bibnamefont {Gygi}},\ }\bibfield  {title} {\enquote {\bibinfo {title} {Optimization algorithm for the generation of oncv pseudopotentials},}\ }\href {\doibase https://doi.org/10.1016/j.cpc.2015.05.011} {\bibfield  {journal} {\bibinfo  {journal} {Comput Phys Commun}\ }\textbf {\bibinfo {volume} {196}},\ \bibinfo {pages} {36--44} (\bibinfo {year} {2015})}\BibitemShut {NoStop}%
\bibitem [{elk()}]{elk}%
  \BibitemOpen
  \href@noop {} {\enquote {\bibinfo {title} {{The Elk Code}},}\ }\bibinfo {howpublished} {\url{http://elk.sourceforge.net/}}\BibitemShut {NoStop}%
\bibitem [{\citenamefont {Matthias}\ \emph {et~al.}(1968{\natexlab{b}})\citenamefont {Matthias}, \citenamefont {Geballe}, \citenamefont {Andres}, \citenamefont {Corenzwit}, \citenamefont {Hull},\ and\ \citenamefont {Maita}}]{doi:10.1126/science.159.3814.530}%
  \BibitemOpen
  \bibfield  {author} {\bibinfo {author} {\bibfnamefont {B.~T.}\ \bibnamefont {Matthias}}, \bibinfo {author} {\bibfnamefont {T.~H.}\ \bibnamefont {Geballe}}, \bibinfo {author} {\bibfnamefont {K.}~\bibnamefont {Andres}}, \bibinfo {author} {\bibfnamefont {E.}~\bibnamefont {Corenzwit}}, \bibinfo {author} {\bibfnamefont {G.~W.}\ \bibnamefont {Hull}}, \ and\ \bibinfo {author} {\bibfnamefont {J.~P.}\ \bibnamefont {Maita}},\ }\bibfield  {title} {\enquote {\bibinfo {title} {Superconductivity and antiferromagnetism in boron-rich lattices},}\ }\href {\doibase 10.1126/science.159.3814.530} {\bibfield  {journal} {\bibinfo  {journal} {Science}\ }\textbf {\bibinfo {volume} {159}},\ \bibinfo {pages} {530--530} (\bibinfo {year} {1968}{\natexlab{b}})}\BibitemShut {NoStop}%
\bibitem [{\citenamefont {Bekaert}\ \emph {et~al.}(2019)\citenamefont {Bekaert}, \citenamefont {Petrov}, \citenamefont {Aperis}, \citenamefont {Oppeneer},\ and\ \citenamefont {Milo\ifmmode \check{s}\else \v{s}\fi{}evi\ifmmode~\acute{c}\else \'{c}\fi{}}}]{PhysRevLett.123.077001}%
  \BibitemOpen
  \bibfield  {author} {\bibinfo {author} {\bibfnamefont {J.}~\bibnamefont {Bekaert}}, \bibinfo {author} {\bibfnamefont {M.}~\bibnamefont {Petrov}}, \bibinfo {author} {\bibfnamefont {A.}~\bibnamefont {Aperis}}, \bibinfo {author} {\bibfnamefont {P.~M.}\ \bibnamefont {Oppeneer}}, \ and\ \bibinfo {author} {\bibfnamefont {M.~V.}\ \bibnamefont {Milo\ifmmode \check{s}\else \v{s}\fi{}evi\ifmmode~\acute{c}\else \'{c}\fi{}}},\ }\bibfield  {title} {\enquote {\bibinfo {title} {{Hydrogen-Induced High-Temperature Superconductivity in Two-Dimensional Materials: The Example of Hydrogenated Monolayer ${\mathrm{MgB}}_{2}$}},}\ }\href {\doibase 10.1103/PhysRevLett.123.077001} {\bibfield  {journal} {\bibinfo  {journal} {Phys. Rev. Lett.}\ }\textbf {\bibinfo {volume} {123}},\ \bibinfo {pages} {077001} (\bibinfo {year} {2019})}\BibitemShut {NoStop}%
\end{thebibliography}%
 \end{document}